\documentclass[12pt,preprint]{aastex}

\newcommand{\kms}{\mbox{km s$^{-1}$}}

\newcommand{\Msun}{\mbox{M$_{\odot}$}}
\newcommand{\Lsun}{\mbox{L$_{\odot}$}}
\newcommand{\arcminspace}{\mbox{$\arcmin$ }}
\newcommand{\arcsecspace}{\mbox{$\arcsec$ }}

\newcommand{\tco}{\mbox{$^{13}$CO$(3\rightarrow2)$}}
\newcommand{\dco}{\mbox{CO$(3\rightarrow2)$}}
\newcommand{\cdo}{\mbox{C$^{18}$O$(3\rightarrow2)$}}
\newcommand{\cdodos}{\mbox{C$^{18}$O$(2\rightarrow1)$}}
\newcommand{\csdos}{\mbox{CS$(2\rightarrow1)$}}
\newcommand{\cssiete}{\mbox{CS$(7\rightarrow 6)$}}
\newcommand{\sio}{\mbox{SiO$(2\rightarrow1)$}}
\newcommand{\hco}{\mbox{HCO$^{+}(1\rightarrow0)$}}
\newcommand{\hcotrece}{\mbox{H$^{13}$CO$^{+}(1\rightarrow 0)$}}
\newcommand{\metanol}{\mbox{CH$_{3}$OH$(2\rightarrow1)$}}
\newcommand{\cstyc}{\mbox{C$^{34}$S$(2\rightarrow1)$}}
\newcommand{\x}{~\mbox{$\times$}}

\slugcomment{Published in ApJ, 715, 18}

\shorttitle{Molecular outflows toward the IRDC G34.43+0.24}
\shortauthors{Sanhueza et al.}

\begin{document}


\title{Molecular outflows within the filamentary infrared dark cloud
G34.43+0.24}

\author{P. Sanhueza\altaffilmark{1}, G. Garay, L. Bronfman, D. Mardones and J. May}
\affil{Departamento de Astronom\'{\i}a, Universidad de Chile,
Casilla 36-D, Santiago, Chile}

\and

\author{M. Saito}
\affil{National Astronomical Observatory of Japan, 2-21-1 Osawa, Mitaka, Tokyo 181-8588, Japan}

\altaffiltext{1}{Current address: Institute for Astrophysical Research, Boston University, Boston, MA 02215, USA; patricio@bu.edu}


\begin{abstract}

We present molecular line observations, made with angular resolutions
of $\sim$20\arcsec, toward the filamentary infrared dark cloud
G34.43+0.24 using the APEX [\dco , \tco, \cdo\ and
\cssiete\ transitions], Nobeyama 45 m 
[\csdos, \sio, \cstyc, \hco , \hcotrece\ and \metanol\ transitions] 
, and SEST [\csdos\ and \cdodos\ transitions] telescopes. We find that  
the spatial distribution of the molecular emission is similar to that
 of the dust continuum emission observed with 11\arcsec\ 
 resolution (Rathborne et al. 2005) showing a filamentary structure
 and four cores. The cores have local thermodynamic equilibrium masses 
ranging from 
 $3.3\times10^2$ -- $1.5\times10^3$ \Msun\ and virial masses from 
$1.1\times10^3$ -- $1.5\times10^3$ \Msun, molecular hydrogen densities
 between $1.8\times10^4$ and $3.9\times10^5$ cm$^{-3}$, 
and column densities $>2.0\times10^{22}$ cm$^{-2}$; values characteristics
of massive star forming cores. The \tco\ profile observed toward the most 
massive core reveals a blue profile indicating that the core is
 undergoing large--scale inward motion with an average infall velocity
 of 1.3 \kms\ and a mass infall rate of $1.8\times10^{-3}$ \Msun\ yr$^{-1}$.
We report the discovery of a molecular outflow toward the northernmost 
core thought to be in a very early stage of evolution.
We also detect the presence of high velocity gas toward each of the 
other three cores, giving support to the hypothesis that the excess 
4.5 $\mu$m emission (``green fuzzies'') detected toward these cores 
is due to shocked gas. The molecular outflows 
 are massive and energetic, with masses ranging from 25 -- 80 \Msun,
 momentum 2.3 -- $6.9\times10^2$ \Msun\ \kms, and kinetic energies
 1.1 -- $3.6\times 10^3$ \Msun\ km$^2$ s$^{-2}$; indicating that they are
 driven by luminous, high-mass young stellar objects. 

\end{abstract}


\section{Introduction}

Infrared dark clouds (IRDCs) are cold ($<$25 K), massive ($\sim$$10^2$ -- 
$10^4$ \Msun), and dense ($>$$10^5$ cm$^{-3}$) molecular clouds with
 high column 
densities ($\sim$$10^{23}$--$10^{25}$ cm$^{-2}$) seen as a dark silhouette 
against the bright mid--infrared background emission  (Perault et al. 1996; 
Egan et al. 1998; Carey et al. 1998, 2000; Hennebelle et al. 2001; Simon et 
al. 2006a, 2006b). A wealth of recent observations show that IRDCs are 
the sites of high-mass stars and star cluster formation (Rathborne et al. 2006, 
2007; Pillai et al. 2006; Jackson et al. 2008; Chambers et al. 2009). 
Dust continuum observations reveal that IRDCs harbor compact cold cores with 
typical masses of $\sim$120 \Msun\ and sizes $<$0.5 pc (Rathborne 
et al. 2006). The masses and sizes of the IRDC cores are similar
 to those of the 
massive hot cores (e.g., Garay \& Lizano 1999), but IRDC cores are much colder. 
Recently, Chambers et al. (2009) investigated a sample of 190 cores found  
toward 38 IRDCs by Rathborne et al. (2006), and classified them as ``active'' 
if they are associated with enhanced 4.5 $\mu$m emission, the so-called green 
fuzzies, and  an embedded 24 $\mu$m source; and ``quiescent'' if they contain 
neither of these indicators. They found 37 active cores and 69 quiescent cores. 
Further, active cores are usually associated with bright ($>$1 Jy) methanol 
and water masers (Pillai et al 2006; Wang et al. 2006; Chambers et al. 2009), 
and have high bolometric luminosities, indicating that they are forming 
high--mass stars (M $>$ 8 \Msun) stars. The quiescent cores are the most 
likely candidates for high--mass starless cores. 

G34.43+0.24 is an IRDC with a filamentary morphology,
extending by $\sim$9\arcmin\ from north to south in equatorial 
projection (9.7 pc at the source distance of 3.7 kpc; Fa\'undez et al. 2004; 
Simon et al. 2006b). This IRDC is in the list of Rathborne et al. (2006) 
and Chambers et al. (2009). It is located roughly 11\arcmin\ north of 
the  ultra-compact (UC) H II complex G34.26+0.15 (Molinari et al. 1996). 
Molecular line observations toward this cloud were first reported by 
Miralles et al. (1994), who from ammonia observations with 1.5\arcmin\ 
resolution found an elongated structure in the N-S direction with a total 
mass of 1000 \Msun. Near the center of the filament lies the IRAS point
source 18507+0121 which has a luminosity of $3.4\times10^4$ \Lsun.
 
Bronfman et al. (1996) detected toward IRAS 18507+0121 strong \csdos\ emission 
with broad line wings, indicating it is associated with a dense massive 
star forming region. Ramesh et al. (1997) observed the IRAS source in the \hco, 
\hcotrece, \csdos\ and \cstyc\  molecular lines, with angular resolutions of  
$\sim$16\arcsec, modeling the observed line profiles as due to emission from 
a collapsing hot core that is hidden behind a cold and dense envelope. From 
observations of 3 mm continuum and H$^{13}$CO$^+$ line emissions, with 
$\sim$5\arcsec\ angular resolution, Shepherd et al. (2004) detected two compact 
molecular cores, separated by 40\arcsec, toward the center of the filament. 
Millimeter continuum observations revealed the presence of four dust
cores (Fa\'undez et al. 2004; Garay et al. 2004; Rathborne et al. 2005),
labeled by the latter authors as MM1, MM2, MM3, and MM4. The MM1 dust
core, which corresponds to the northern compact molecular core of
Shepherd et al. (2004), contains a deeply embedded luminous object
surrounded by several hundred solar masses of warm gas and dust.
Based on the weak 6 cm continuum emission and the lack of detection at NIR 
wavelenghts, Shepherd et al. (2004) suggested that the embedded object appears 
to be a massive B2 protostar in an early stage of evolution. The MM2 dust 
core, which corresponds to the southern compact molecular core of Shepherd 
et al. (2004), is associated with IRAS 18507+0121; with a NIR cluster of young 
stars with a central B0.5 star (Shepherd et al. 2004); with an UC H II region 
(Miralles et al. 1994; Molinari et al. 1998); with a variable H$_2$O maser 
(Miralles et al. 1994); and with CH$_3$OH maser emission
 (Szymczak et al. 2000). 
The MM3 dust core is located close to the northern edge of the filament, 
$\sim$3.5\arcminspace north of MM2. Garay et al. (2004) suggested that MM3 
corresponds to a massive and dense cold core that will eventually collapse 
to form a high-mass star. The MM4 core is located $\sim$30\arcsecspace south
of MM2 (Rathborne et al.  2005). Table~\ref{tbl-1.2mm} lists the measured 
1.2 mm dust continuum emission properties as measured by Rathborne et al. 
(2006). Rathborne et al. (2005) and Chambers et al (2009) reported an excess 
4.5 $\mu$m emission toward MM1, MM3, and MM4 cores, which could be produced
by either ionized gas and/or shocked gas and acts therefore as a signature of
current star formation (MM2 is classified by the last authors as a ``red'' 
core, namely with bright 8 $\mu$m emission). They also showed that the four 
MM cores are associated with 24 $\mu$m point sources (emission from warm dust), 
which is an indicator of accretion onto an embedded protostar. 
Water maser emission, a well--known tracer of low--mass and high--mass star 
formation, has been detected toward all four MM cores (Wang et al. 2006; 
Chambers et al. 2009). Chambers et al. (2009) detected Class I CH$_3$OH masers,
another well--know signpost of star formation toward MM1, MM2, and 
MM3 cores. Rathborne et al. (2006) identified nine 1.2 mm continuum cores with 
masses ranging from 80 to 1300 \Msun\ toward G34.43+0.24.
 
Shepherd et al. (2007) observed the central region of the filament,
 encompassing 
the MM1 and MM2 cores, at millimeter wavelengths, in CO$(1\rightarrow0)$, 
$^{13}$CO$(1\rightarrow0)$, and C$^{18}$O$(1\rightarrow0)$ at
 $\sim4$\arcsec\ angular resolution and at near infrared wavelengths
with the {\it Spitzer} Space Telescope. They discovered five massive
outflows, three of them associated with the MM2 core and the remaining
two associated with the MM1 core. From the {\it Spitzer} data they
identified 31 young stellar objects in the filament, with a combined
mass of $\sim$127 \Msun, and additional 22 sources that they believe
could be cluster members based on the presence of strong
24 $\mu$m emission. Cortes et al. (2008) observed MM1 and MM2 in polarized 
thermal dust emission at 3 mm and CO$(1\rightarrow0)$ line emission  
with BIMA array. Their results suggest that there is a magnetic field 
othogonal to the axis formed by the MM1 and MM2 cores.   
Recently, Rathborne et al. (2008) carried out high angular submillimeter 
observations toward the MM1 core using the Submillimeter Array. 
They determined, using 
continuum emission images, that MM1 remains unresolved with a size of 
$\sim$0.03 pc and a mass of 29 \Msun. Their molecular line spectrum 
shows the presence of several complex molecules, suggesting that MM1 is 
a hot core. They also found an extended \tco\ structure which may be    
evidence of a rotating envelope surrounding 
the central high--mass star, perpendicular to one of the outflows that 
Shepherd et al. (2007) detected toward MM1. 

In this work we report observations of 11 molecular transitions 
using the APEX 12-m, SEST 15-m, and Nobeyama 45-m telescopes toward 
the G34.43+0.24 cloud. Since different molecular species trace
different density and kinematics regimes, the capability of
observing multiple spectral lines is particularly advantageous to
determine the morphology and kinematics of star forming cores. Due
to the high abundance of CO, the \dco\ line is an ideal tracer of
the low intensity, high--velocity wing emission. The \tco\ line,
being more optically thin than \dco, traces, in addition to
the ambient gas, the low--velocity gas of the molecular outflow.
The \sio\ line is usually associated with the presence of outflows
since its abundance is highly enhanced by shocks. Methanol transitions are 
usually used as tracers of dense gas, temperature probes and tracers of 
shocks, since this molecule is evaporated from grain mantles. 
The \hco\ and
\csdos\ lines are high--density tracers and thus are good probes
of the dense envelopes surrounding protostars. The \cdo, \cdodos,
\hcotrece, \cssiete\ and \cstyc\ molecular transitions probe the
inner regions of the cores. The last two species are weakly
influenced by infall and outflow motions, and are used as reference
to estimate the central ambient cloud velocity.

The main goals of these multi--line observations are to determine
the morphology and kinematics of the gas within the molecular cores
in the G34.43+0.24 filamentary IRDC, in particular to 
investigate the presence of outflowing gas, in order to increase our  
knowledge of the evolutionary state of the cores and to achieve a better 
understanding of the overall process of massive star formation along 
the filament. This is the first survey of several molecular lines mapping 
portions and the whole cloud. It gives us, for the first time, a  
comprehensive view of the molecular gas distribution at the scale 
of parsecs.

\section{Observations}

The observations were made using the 12-m APEX\footnote{Atacama
Pathfinder EXperiment. APEX is a collaboration  between the
 Max-Planck-Institut fur Radioastronomie, the European Southern
 Observatory and the Onsala Space Observatory.} telescope,
 the 15-m SEST telescope, both located in Chile, and the 45-m telescope
 of the Nobeyama Radio Observatory (NRO) located in Japan. The
 molecules, transitions, frequencies and observing parameters are
 listed in Table~\ref{tbl-obsparam}.

\subsection{APEX Telescope}

With the APEX telescope we observed the \dco, \tco, \cdo\ and
 \cssiete\ lines. The observations were carried out during three 
epochs: 2005 July, 2005 October, and 2006 May-July. The telescope
 beam size at the frequency of the observed lines ranged from 17 to
 19\arcsec\ (FWHM). We used the heterodyne receiver (APEX-2a),
 which provided a channel separation of 122 kHz and a total
 bandwidth of 1 GHz with 8192 spectral channels. 

In \dco\ and \tco, we mapped the emission within a
 100\arcsec$\times$680\arcsecspace region which cover a spatial
 extent similar to that exhibited by the 1.2 mm dust continuum
 emission map (see Garay et al. 2004). The spacing between adjacent points 
on a regular grid was 20\arcsecspace and the position-switching mode was used. 
The on-source integration
 time per map position was typically $\sim$40 s in both molecules. 
 System temperatures were in the range 110-340 K in \dco\ and 
 200-300 K in \tco, resulting in an rms noise per channel, in antenna 
temperature, 
 of typically 0.27 K and 0.37 K, respectively. In \cdo\ we mapped
 the emission within a 80\arcsec$\times$100\arcsecspace zone,
 covering the central region of the filament also known as the
 IRAS 18507+0121 region. System temperatures were typically
 $\sim$190 K. The on-source integration time per map position was
$\sim$50 s resulting an rms noise per channel of typically 0.29 K in antenna 
temperature. In \cssiete\ we mapped the emission within a
40\arcsec$\times$120\arcsecspace region. System temperatures were 
typically $\sim$200 K. On-source integration time was $\sim$80 s
 resulting an rms noise per channel of typically 0.18 K in antenna temperature.

The molecular line data were reduced according to the standard
 procedure using CLASS, a GILDAS working group software.

\subsection{NRO 45 m Telescope}

With the NRO 45 m telescope we observed the \csdos, \sio, \cstyc,
\hco, \hcotrece\ and \metanol\ lines toward the central region of
the filament. The observations were undertaken during 1996 February-April. 
The telescope beam size at the frequencies of the observed
 lines is 15-18\arcsec. We used an acousto-optical spectrometer with
 37 kHz resolution covering a total bandwidth of 40 MHz.

The emission was mapped in all lines within a
 80\arcsec$\times$140\arcsecspace region. The spacing between adjacent 
points on a regular grid was 20\arcsec, using the position-switching mode.
 System temperatures were in the
 range between 160 and 390 K. On-source integration times per map
 position ranged from 300 to 600 s resulting in rms noise 
of typically 0.08 K per channel in antenna temperature in all lines. 
First--order polynomials were fitted in order to estimate and remove 
the baselines. Part of this data were used by Ramesh, Bronfman \& 
Deguchi (1997) to model the kinematics at the peak position 
of IRAS 18507+0121.

\subsection{SEST Telescope}

With the SEST telescope we observed the \csdos\ and
\cdodos\ lines toward the northern part of the filament (MM3 core).
The telescope beam size is 24\arcsec\ at the frequency of the
\cdodos\ line and 51\arcsec\ at the frequency of the \csdos\ line.
The observations were carried out in 2003 March. We used the
 high-resolution acousto-optical spectrometers, which provided a
 channel separation of 43 kHz and a total bandwidth of 43 MHz. We
 observed, with 30\arcsec\ spacing between adjacent 
points on a regular grid, 25 positions within a
 150\arcsec$\times$150\arcsec\ region.  The on-source integration
 times per map position were typically $\sim$130 s for CS and
 $\sim$70 s for C$^{18}$O. System temperatures were in the range 
230-270 K for CS and 400-540 K for C$^{18}$O, resulting in an rms
 noise, in antenna temperature, of typically 0.07 K and 0.18 K per channel,
 respectively.

\section{Results}

\subsection{Filament and Dense Cores}

\begin{figure}

\begin{center}

\includegraphics[angle=0,scale=0.85]{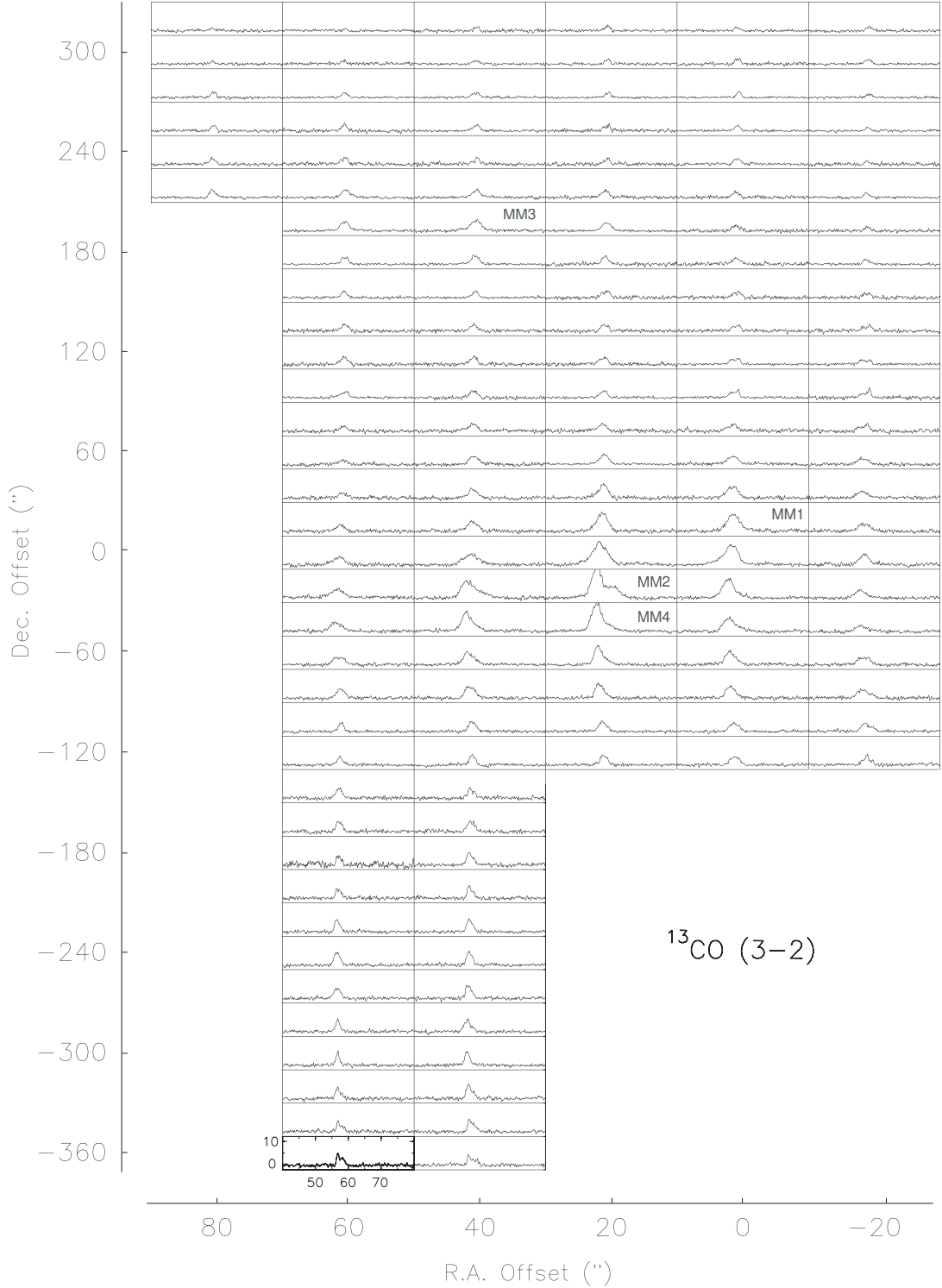}

\end{center}

\caption{Observed spectra of the \tco\ line emission toward G34.43+0.24,
taken with the APEX telescope. The angular separation between panels is
20\arcsec.  Offsets are from the reference position at $\alpha = 18^{h}
53^{m} 17.^{s}40$ and $\delta = 01^{\circ} 24 \arcmin 55 \arcsec $ (J2000).
In each box the velocity scale ranges from 40 to 80 \kms. The antenna
temperature scale is from -2 to 12 K. }

\label{fig-13co32-spectra}
\end{figure}

Figure~\ref{fig-13co32-spectra} shows the \tco\ spectra observed with
 APEX along the whole filamentary structure detected in the 1.2 mm
 dust continuum observations of Garay et al. (2004; see their Figure 4).
 Emission in the \tco\ line was detected across the whole mapped region.

\begin{figure}
\begin{center}
\includegraphics[angle=0,scale=0.8]{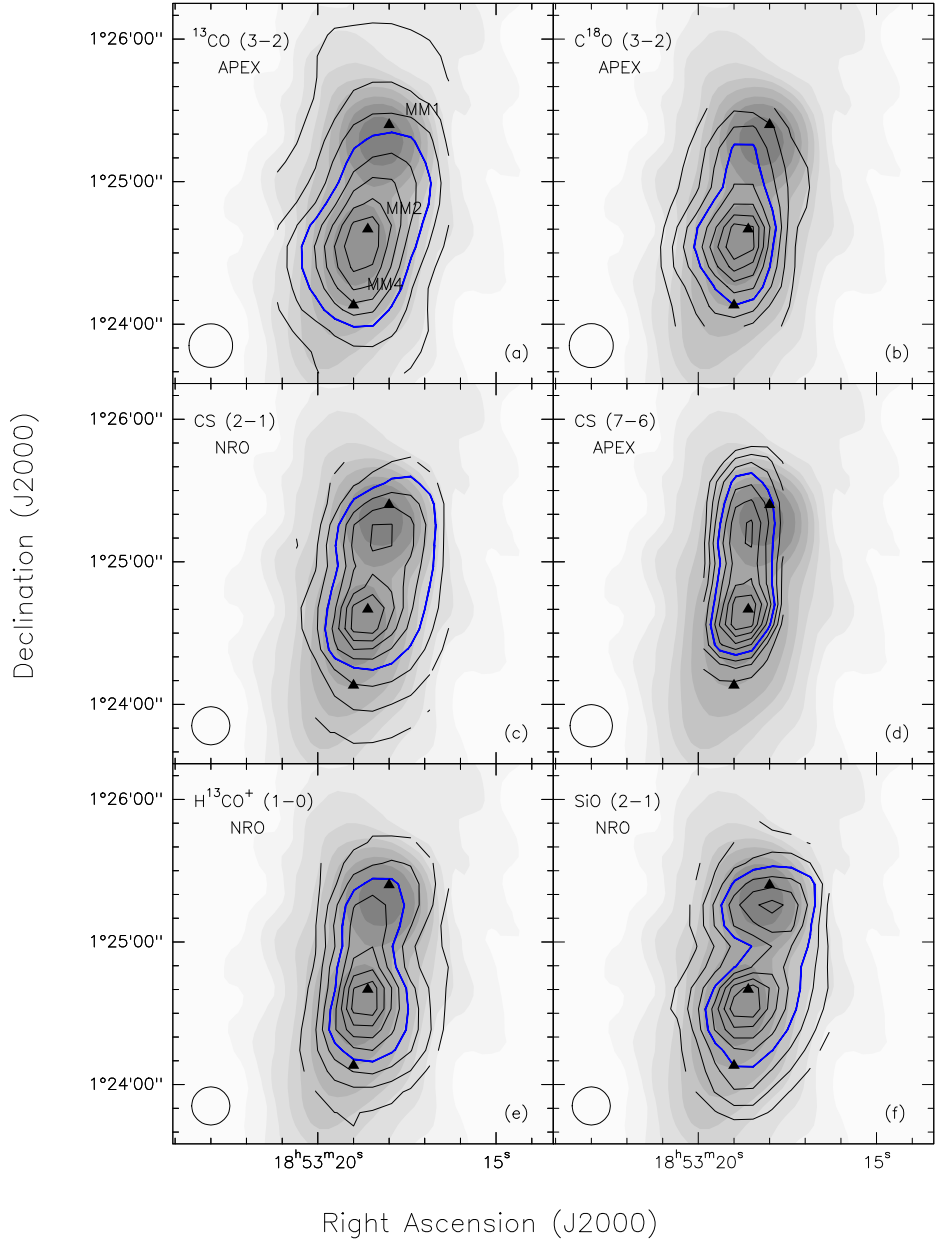}
\end{center}
\caption{ {\footnotesize Contour maps of velocity-integrated
ambient molecular emission toward the central region of G34.43+0.24,
overlaid on a 1.2 mm dust continuum emission map (gray scale).
The range of integration in $v_{lsr}$ is from 52.9 to 61.1 \kms.
The FWHM beam is shown in the bottom left corner.
Contour levels are from 20\% to 90\% (in steps of 10\%) of the peak
integrated intensity of 49.9 K \kms\ for \tco\ ({\it a}); 7.7 K \kms\
for \cssiete\ ({\it b});  16.9 K \kms\ for \csdos\ ({\it c});  3.4 K \kms\
for \sio\ ({\it d}); 3.7 K \kms\ for \hcotrece\ ({\it e}); and 14.3 K \kms\
for \cdo\ ({\it f}). The blue lines indicate
 the 50\% contour level. The gray scale levels are 1, 2, 3, 5, 7, 10, 15,
 20 and 30 $\times$ 0.12 Jy beam$^{-1}$. The triangles mark the peak
positions of the  MM1, MM2 and MM4 dust cores. } }
\label{fig-ambientcentral}
\end{figure}

Figure~\ref{fig-ambientcentral} shows contour maps of the velocity
 integrated ambient gas emission (velocity range from 52.9 to 61.1
 \kms, determined by inspection of the line profiles) in the \tco, 
\cssiete, \csdos, \sio, \hcotrece\ and \cdo\
 lines observed toward the central $\sim$1.5\arcmin$\times$3\arcmin\
 region of the filament. The maps are overlaid on a grey scale image
 of the 1.2 mm dust continuum emission observed with SIMBA/SEST by 
Garay et al. (2004).  
We find that the morphology of the molecular line emission resembles
 that of the dust continuum emission, showing two components: a
 northern component associated with the MM1 core and a southern
 component associated with the MM2 dust core. There are, however,
 some differences in the morphology of the emission in the different
 molecular lines. These are likely due to chemistry and/or optical
 depth effects and/or differences in the regimes of density and
 temperature traced by the different lines. In particular, the
 emission in the \cssiete\ line is less extended than in the other
 transitions and runs nearly parallel to the north-south direction.

\begin{figure}
\begin{center}
\includegraphics[angle=0,scale=0.8]{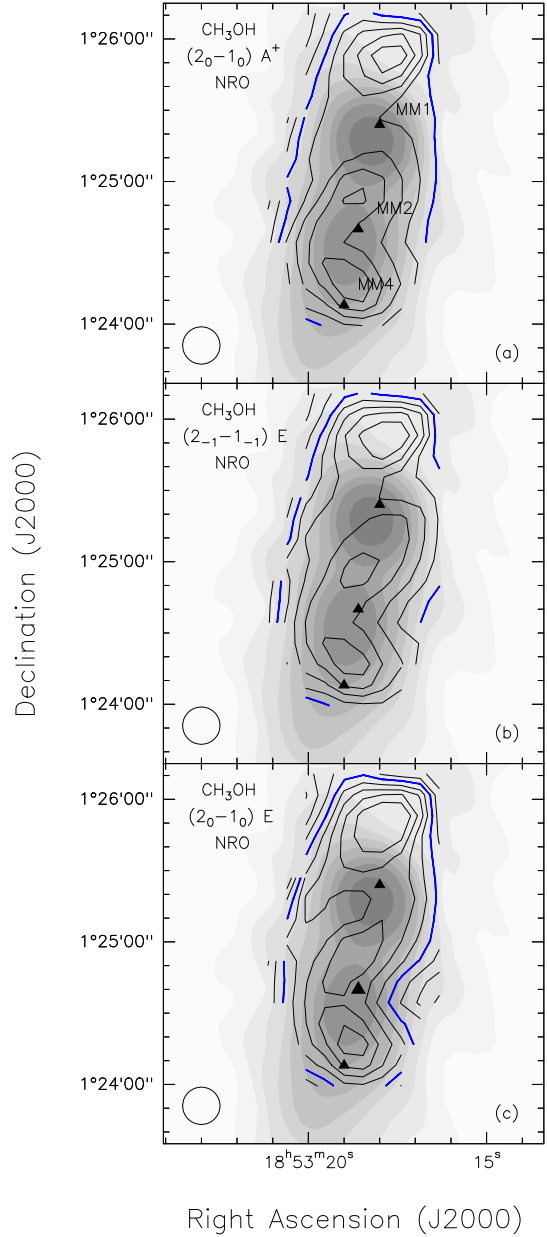}
\end{center}
\caption{ {\footnotesize Contour maps of velocity-integrated
ambient CH$_3$OH emission toward the central region of G34.43+0.24,
overlaid on a 1.2 mm dust continuum emission map (gray scale).
The FWHM beam is shown in the bottom left corner.
Contour levels are from 20\% to 90\% (in steps of 10\%) of the peak
integrated intensity of 5.1 K \kms\ for CH$_3$OH(2$_0\rightarrow1_0$) A+
 ({\it a}); 4.1 K \kms\ for CH$_3$OH(2$_{-1}\rightarrow1_{-1}$) E ({\it b});
and 1.1 K \kms\ for CH$_3$OH(2$_{0}\rightarrow1_{0}$) E ({\it c}). The blue 
lines indicate the 50\% contour level. The gray scale levels are 
1, 2, 3, 5, 7, 10, 15, 20 and 30 $\times$ 0.12 Jy beam$^{-1}$. The 
triangles mark the peak positions of the  MM1, MM2 and MM4 dust cores. } }
\label{CH3OH_mapa_pos_pos}
\end{figure}

Figure~\ref{CH3OH_mapa_pos_pos} presents contour maps of the velocity 
integrated ambient gas emission (velocity range from 55.8 to 60.6 \kms)
in three methanol lines [CH$_3$OH(2$_0\rightarrow1_0$) A+,  
CH$_3$OH(2$_{-1}\rightarrow1_{-1}$) E, 
CH$_3$OH(2$_{0}\rightarrow1_{0}$) E] observed toward the central
 region of G34.43+0.24. The maps are overlaid on a gray scale image
 of the 1.2 mm dust continuum emission observed with SIMBA/SEST. The
 morphology of the emission in the three methanol lines is all
 similar. They are, however, different from the morphology exhibited
 by the dust continuum emission and the rest of the molecular lines.
 The methanol peaks do not coincide with the dust peaks, on the 
contrary they seem to be slightly anticorrelated. This could be 
due to depletion of methanol within the central regions of the dust 
cores.
 
\begin{figure}
\begin{center}
\includegraphics[angle=0,scale=0.85]{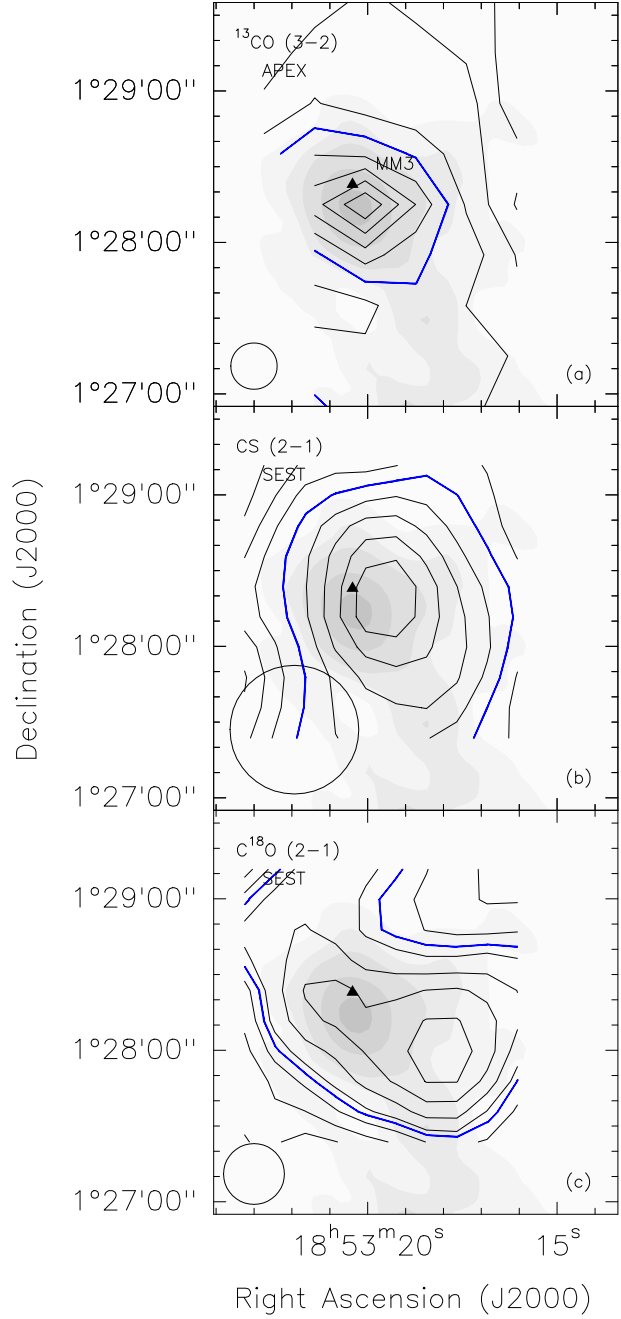}
\end{center}
\caption{ {\footnotesize Contour maps of velocity-integrated
molecular line emission toward the MM3 core (Garay et al. 2004), 
overlaid on a 1.2 mm dust continuum emission map (gray scale).
The range of integration in $v_{lsr}$ is from 56.1 to 61.5 \kms. Contour
levels are from 20\% to 90\% (in steps of 10\%) of the peak integrated
intensity of 12.5 K \kms\ for \tco\ ({\it a}) and 3.7 K \kms\ for
\csdos\ ({\it b}); and from 3 to 21 $\sigma$ in steps 3 $\sigma$,
with $\sigma$ = 0.14 K \kms, for \cdodos\ ({\it c}).
The gray scale levels are 1, 2, 3, 5, 7, 10, 15,
 20 and 30 $\times$ 0.12 Jy beam$^{-1}$. The blue lines indicate
 the 50\% contour level. The triangle marks
the peak position of the MM3 core.
The FWHM beam is shown in the bottom left corner.}}
\label{fig-ambientmm3}

\end{figure}

Figure~\ref{fig-ambientmm3} shows contour maps of velocity-integrated
ambient gas emission (velocity range from 56.1 to 61.5 \kms)
in the \tco, \csdos\ and \cdodos\ lines observed toward the MM3 core,
overlaid on a gray scale image of the 1.2 mm dust continuum emission.
Whereas the morphology of the emission in the \tco\ line is similar to 
that of the dust continuum emission, that in the \cdodos\ line is 
somewhat different, showing a peak that is displaced from the
peak of the dust emission.

Table~\ref{tbl-sizes} gives the observed parameters of the molecular
 cores. The angular sizes, determined from the contour maps,
 correspond to the observed semi-major and semi-minor axes. 
The line center velocity, line width and velocity-integrated brightness 
temperature were determined from Gaussian fits to the composite spectra
 of the ambient cloud emission within each core. The cores are assumed
 to be at a distance of 3.7 kpc, which corresponds to the kinematical
 distance of IRAS 18507+0121 reported by Fa\'undez et al. (2004) and  
the kinematical distance of the IRDC given by Simon et al. (2006b). 

\subsection{Spectra Observed Toward the Dense Cores}

There are notable differences between the line profiles observed
 in different molecular species, for different positions. In what
 follows we describe the spectra observed toward three selected
 positions (0\arcsec,~20\arcsec), (20\arcsec,~-20\arcsec) and
 (40\arcsec,~200\arcsec) associated with, respectively, the MM1,
 MM2 and MM3 dust peaks.

\begin{figure}
\begin{center}
\includegraphics[angle=0,scale=0.9]{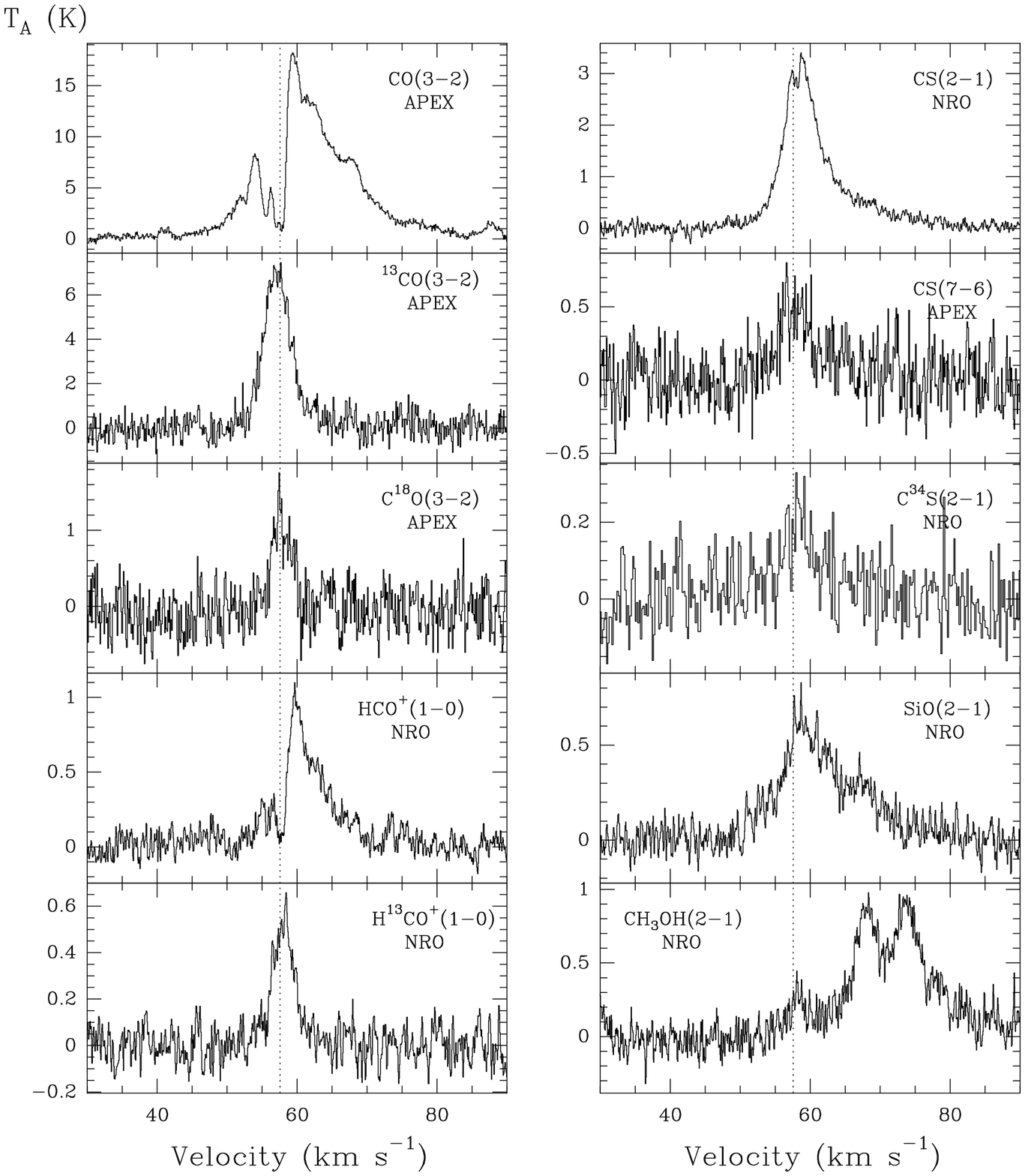}
\end{center}
\caption{Molecular spectra observed toward the peak position of the millimeter 
continuum of MM1 core. Transitions
and telescopes are given in the top corner of the spectra. The vertical
dotted line indicates the systemic velocity of the ambient gas of 57.6 \kms\
(Miralles, Rodr\'iguez \& Scalise 1994).}
\label{fig-specpeakmm1}
\end{figure}

Figure~\ref{fig-specpeakmm1} shows the spectra observed toward the 
(0\arcsec,~20\arcsec) position, which corresponds to the MM1  
core. The \dco\ profile shows, in addition to considerable 
self-absorption, the presence of both blueshifted and redshifted
 high velocity emission. The strong \dco\ wing emission at
 redshifted velocities is also clearly seen in the \sio, \csdos,
 and \hco\ spectra.
 
\begin{figure}
\includegraphics[angle=0,scale=0.9]{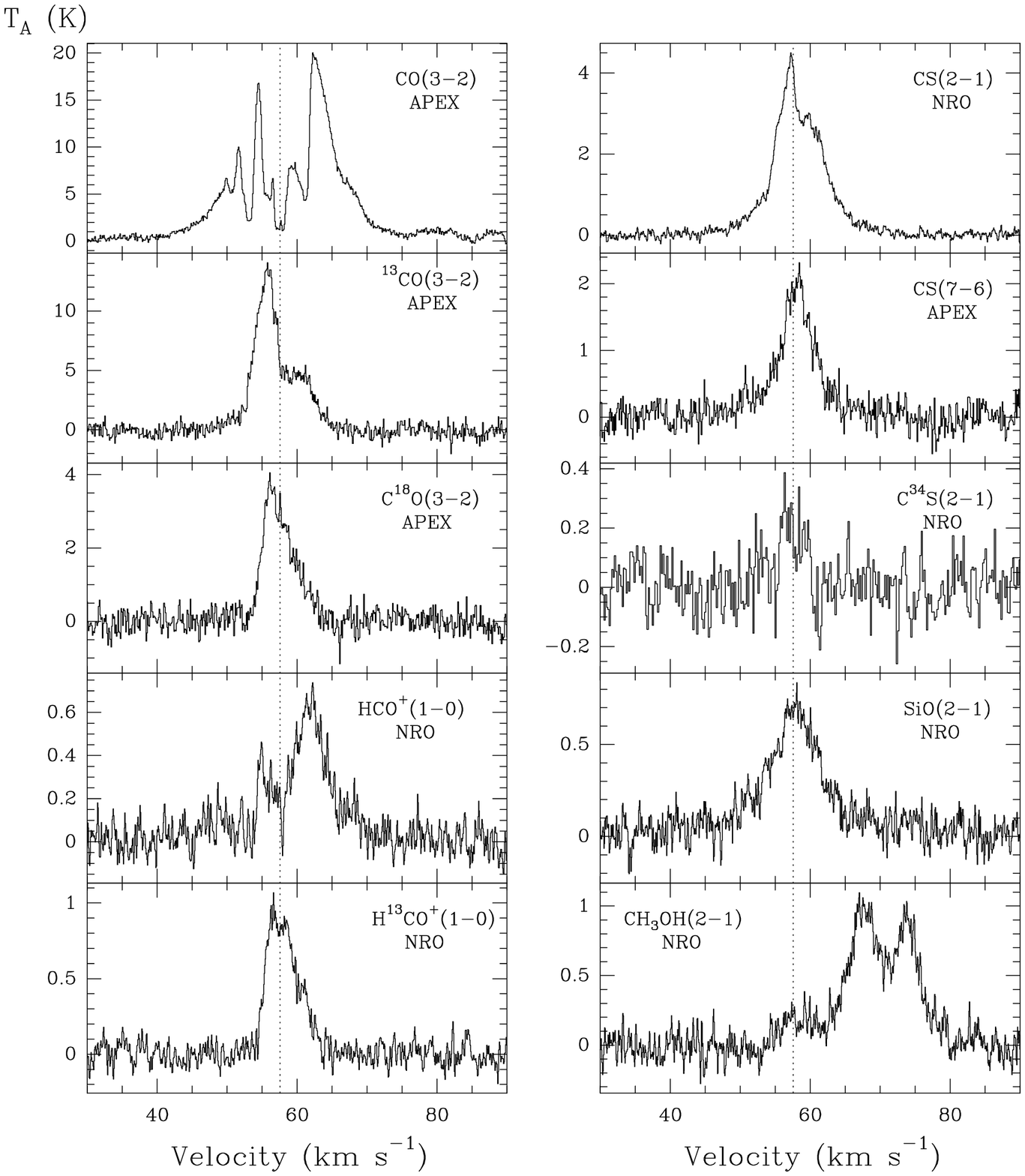}
\caption{Molecular spectra observed toward the peak position of the millimeter 
continuum of MM2 core (IRAS 18507+0121).
Transitions and telescopes are given in the top corner of the spectra. The
vertical dotted line indicates the systemic velocity of the ambient gas
 of 57.6 \kms\ (Miralles, Rodr\'iguez \& Scalise 1994).}
\label{fig-specpeakmm2}
\end{figure}

Figure~\ref{fig-specpeakmm2} shows the spectra observed toward the 
(20\arcsec,$-$20\arcsec) position (MM2 core). 
The \dco\ profile shows considerable self-absorption,
 and the presence of high velocity emission at both blueshifted and
 redshifted velocities.
The \tco\ line profile exhibits a strong blueshifted feature, with a
 peak at the velocity of 55.8 \kms, and a flat redshifted shoulder. 
The \cdo\ line shows strong emission toward blueshifted velocities,
 with a peak at the velocity of 56.1 \kms, and a skewed profile
 toward redshifted velocities. A similar profile is seen in 
the \hcotrece\ line.
The \csdos\ and \hco\ lines show double-peaked profiles; the former
 exhibits a strong blue peak and a weaker red peak whereas the later
 shows an opposite asymmetry, namely a strong redshifted peak and a
 weaker blueshifted peak. The \cssiete\ and \cstyc\ lines exhibit
 symmetrical profiles, with peak line center velocites, determined
 from Gaussian fits, of 58.1 $\pm$ 0.1 \kms\ and 57.4 $\pm$ 0.2 \kms,
 respectively. 

Ramesh et al. (1997) modeled the observed \csdos, \cstyc, \hco\ and
 \hcotrece\ profiles toward MM2 as being produced by a collapsing hot
 core hidden behind a cold intervening screen or envelope. We have
 modeled here the line profiles of \tco, \cdo\ and \cssiete\ using
 the simple model of spectral-line profiles from contracting clouds
 of Myers et al. (1996). We concluded that the MM2 core is undergoing
 large-scale inward motions with a infall velocity of $\sim$1.3 
\kms\ (see section~\ref{infall}). 

\begin{figure}
\begin{center}
\includegraphics[angle=0,scale=0.9]{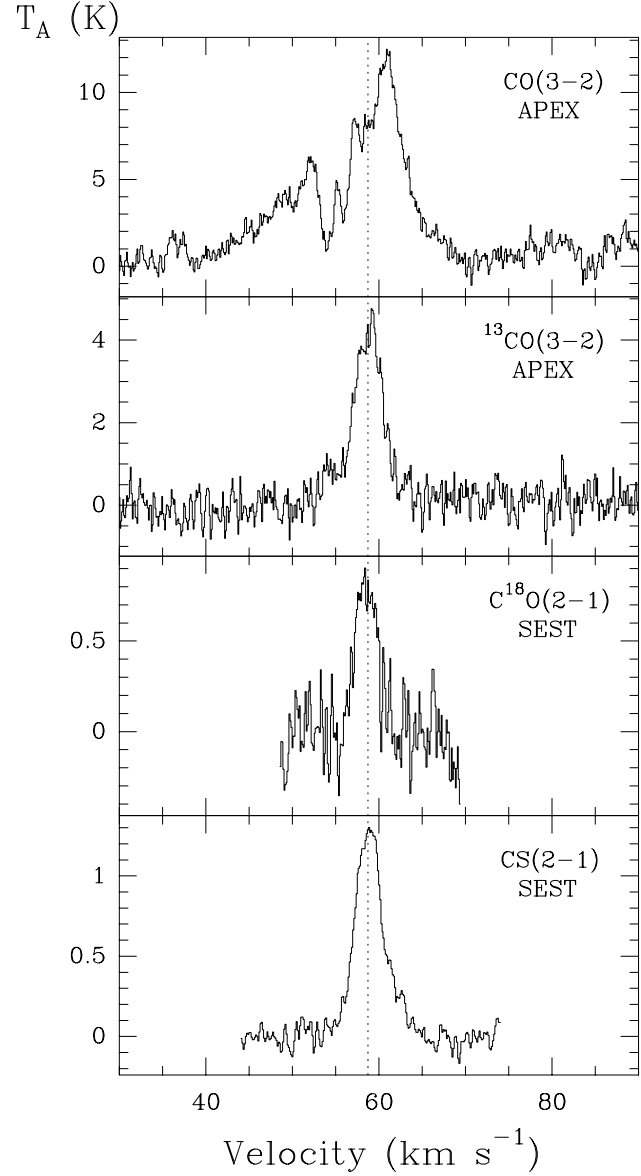}
\end{center}
\caption{Molecular spectra observed toward the peak position of the millimeter 
continuum of MM3 core. Transitions and
telescopes are given in the top right corner of the spectra. The vertical
dotted line indicates the systemic velocity of the ambient gas of 58.7 \kms\
(estimated from the optically thin \tco , \cdodos\ and \csdos\ lines).}
\label{fig-specpeakmm3}
\end{figure}

Figure~\ref{fig-specpeakmm3} shows the spectra observed toward the
(40\arcsec,~200\arcsec) position, which corresponds to the MM3
 core. The \dco\ spectrum shows, in addition to considerable
 self-absorption, strong blueshifted wing emission up to a velocity
 of 42.2 \kms. The \tco\ spectrum exhibits a strong Gaussian line,
 with a center velocity of 58.5 $\pm$ 0.1 \kms, and blueshifted wing
 emission up to a velocity of 53.6 \kms. These observations strongly
 indicate the presence of an energy source within the MM3 core. The
 \cdodos\ and \csdos\ spectra exhibit nearly Gaussian lines profiles
 with peak line center velocities of 58.9 $\pm$ 0.1 and 58.8 $\pm$ 0.1
 \kms, respectively.

\subsection{Molecular Outflows}

\subsubsection{Central Region}

\begin{figure}
\begin{center}
\includegraphics[angle=0,scale=0.68]{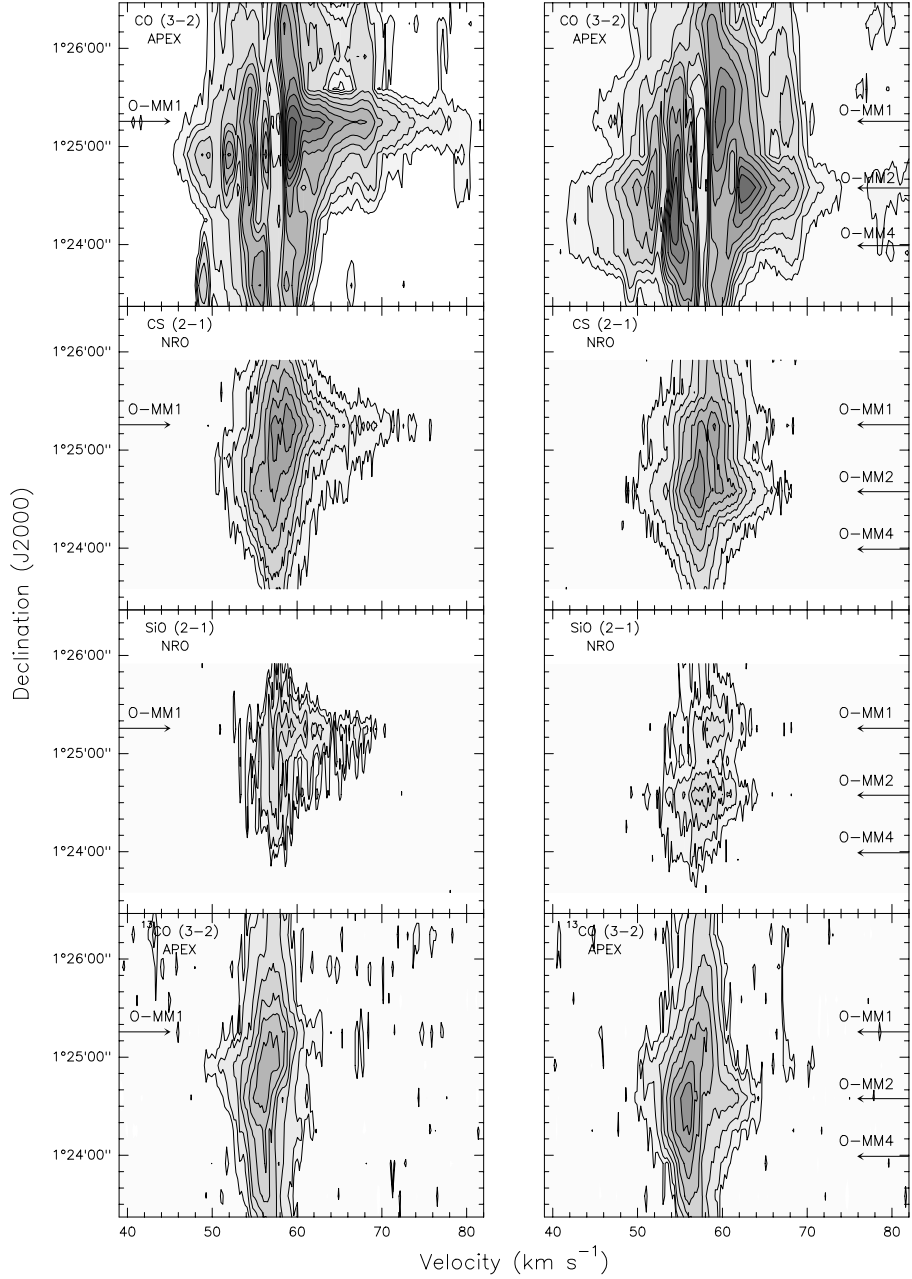}
\end{center}

\caption{{\footnotesize Position-velocity diagrams along lines in the north-south direction 
(i.e., parallel to the filament). Transitions and telescopes are indicated in 
the top left corner of each panel.
{\it Left}: Cut passing through MM1 or position (0\arcsec,~20\arcsec).
Contour levels are 
3, 9, 15, 23, 31, 41, 51, and 61 $\times$ 0.27 K for \dco; 
3, 6, 9, 12, 17, 22, 28, and 34 $\times$ 0.08 K for \csdos; 
3, 4, 6, and 8 $\times$ 0.07 K for \sio; and  
1.5, 3, 6, 9, 12, 15, and 19 $\times$ 0.37 K for \tco.
{\it Right}: Cut passing through MM2 or position (20\arcsec,~0\arcsec). 
Contour levels are 
3, 8, 13, 18, 28, 38, 48, and 58 $\times$ 0.27 K for \dco;
3, 6, 11, 16, 21, 28, 35, and 42 $\times$ 0.08 K for \csdos; 
3, 5, 7, and 9 $\times$ 0.07 K for \sio; and
1.5, 3, 6, 10, 14, 19, 25, and 31 $\times$ 0.37 K 
for \tco.} }

\label{fig-wingpos-vel_MM1andMM2}
\end{figure}

Figure~\ref{fig-wingpos-vel_MM1andMM2} shows position-velocity
 diagrams of the \dco, \csdos, \sio, and \tco\ emission, toward the
 central region of G34.43+0.24, along two cuts in the north-south
 direction passing through the peak position of the MM1 (left panel)
 and MM2 (right panel) cores. Strikingly seen in this Figure is the
 presence of high velocity gas associated with the MM1 and MM2 cores,
 most likely due to molecular outflows emanating from both cores.
Particularly notable is the detection of emission in the silicon
 monoxide line, which is a powerful tracer of the presence of strong
 shocks. The abundance of SiO molecules is highly enhanced, with
 respect to ambient abundances, in molecular outflows as a result
 of destruction of dust grains by shocks, giving rise to the
 injection into the gas phase of Si atoms and/or Si-bearing species.
 Once silicon is injected into the gas phase, chemical models based
 on ion-molecule reactions predict a large abundance of SiO 
molecules (Turner \& Dalgarno 1977; Hartquist et al. 1980).

In the cut passing through the peak of MM1, the most prominent
 outflow emission is seen redshifted with respect to the ambient gas
 velocity, extending up to velocities of 80.7 \kms\ in \dco\ and
 70.2 \kms\ in \csdos\ and \sio. Weak or no outflow emission is
 detected in the \tco\ line, indicating low optical depths.
In the cut passing through the peak of MM2, both blueshifted and
redshifted high velocity emission are clearly seen emanating from the
MM2 core. In \dco\ the blueshifted and redshifted emissions extend
 up to velocities of 42.0 and 74.2 \kms, respectively. Also seen in
 this cut is wing emission emanating from the MM1 core as well as
 from the MM4 core, suggesting that MM4 also contains an outflow.

\begin{figure}

\begin{center}

\includegraphics[angle=0,scale=0.82]{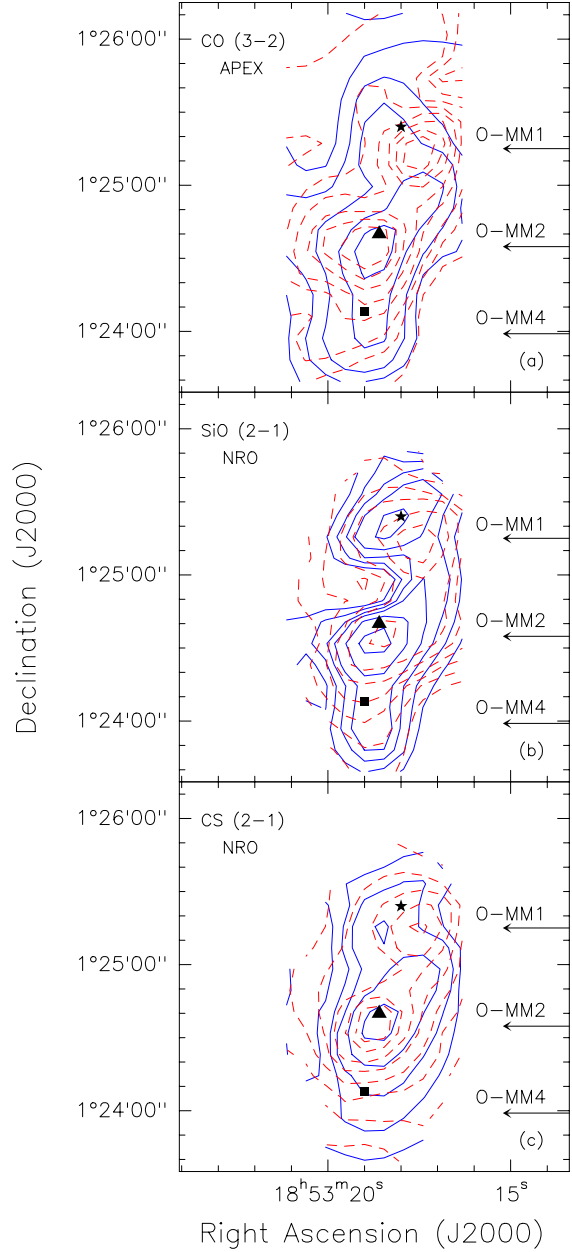}

\end{center}

\caption{ { \footnotesize
Contour maps of velocity-integrated wing emission toward the central
region of G34.43+0.24, in different molecular lines. Transitions and
telescopes are indicated in the top left corner of each panel. 
The arrows in the panels indicate where are the outflows corresponding 
to each core.
 Contour levels for the blueshifted emission are
 20, 35, 50, 80, and 110 $\sigma$ for \dco;
3, 5, 7, 9, 12, and 16 $\sigma$ for \sio; 
and 
 3, 9, 15, 25, and 40 $\sigma$ for \csdos.
 For the redshifted emission, these are
 20, 35, 50, 80, 110, 150, 190, and 230 $\sigma$ for \dco;
 3, 5, 7, 9, 12, 16, 24, and 32 $\sigma$ for \sio;
and   
3, 9, 15, 25, 40, 60, 85, and 110 $\sigma$ for \csdos.
The $\sigma$ values are: 0.30 K \kms\ for the CO line, and 0.04 K \kms\ for
 the SiO and CS lines.}}
\label{fig-wingpos-pos_MM1andMM2}
\end{figure}

Maps of the velocity integrated line wing emission, in the \dco,
 \sio\ and \csdos\ lines toward the central 
1.5\arcmin$\times$3\arcmin\ region of G34.43+0.24, including the 
MM1, MM2 and MM4 dust cores, are shown in 
 Figure~\ref{fig-wingpos-pos_MM1andMM2}. In \sio\ and \csdos, the
 range of velocity integration of the blueshifted 
emission (blue contours or solid lines) is $49.0<v_{lsr}<52.9$ \kms,
 whereas that of the redshifted emission (red contours or dashed
 lines) is $61.1< v_{lsr}<65.2$ \kms. In \dco\ the range of velocity
 integration of the blueshifted and redshifted emissions are,
 respectively, $42.0<v_{lsr}<53.0$ \kms\ and $62.0< v_{lsr}<74.0$ \kms.
In Fig.~\ref{fig-wingpos-pos_MM1andMM2} we identify two, 
possibly three, outflows associated with the dust cores within
 the region considered. The northernmost outflow (labeled O-MM1)
 is associated with the MM1 core. It is roughly oriented in the
 NE-SW direction, with the peaks of the redshifted and blueshifted
 emission being located, respectively, SW and NE of the 3 mm
 continuum source detected by Shepherd et al. (2004; marked with a 
star). There is, however, a large spatial overlap between the
 blueshifted and redshifted emissions.

The outflow associated with the MM2 core, (labeled O-MM2), shows a
 complex morphology with the blueshifted and redshifted emission
 exhibiting a considerable spatial overlap. The peak position of
 the redshifted and blueshifted emissions are located
 $\sim$7\arcsec\ south of the UC HII region (marked with a
 triangle) and are coincident within the errors. This suggest that the
 symmetry axis of the outflow is close to the line of sight.

Outflow-MM4 is associated with the MM4 dust core (marked with a
 square). The blueshifted and redshifted emissions exhibit a large
 overlap in their spatial distribution, suggesting that the flow is
 unresolved or that its axis of symmetry is close to the line of sight.

\subsubsection{Northern Region}

\begin{figure}

\begin{center}

\includegraphics[angle=0,scale=0.9]{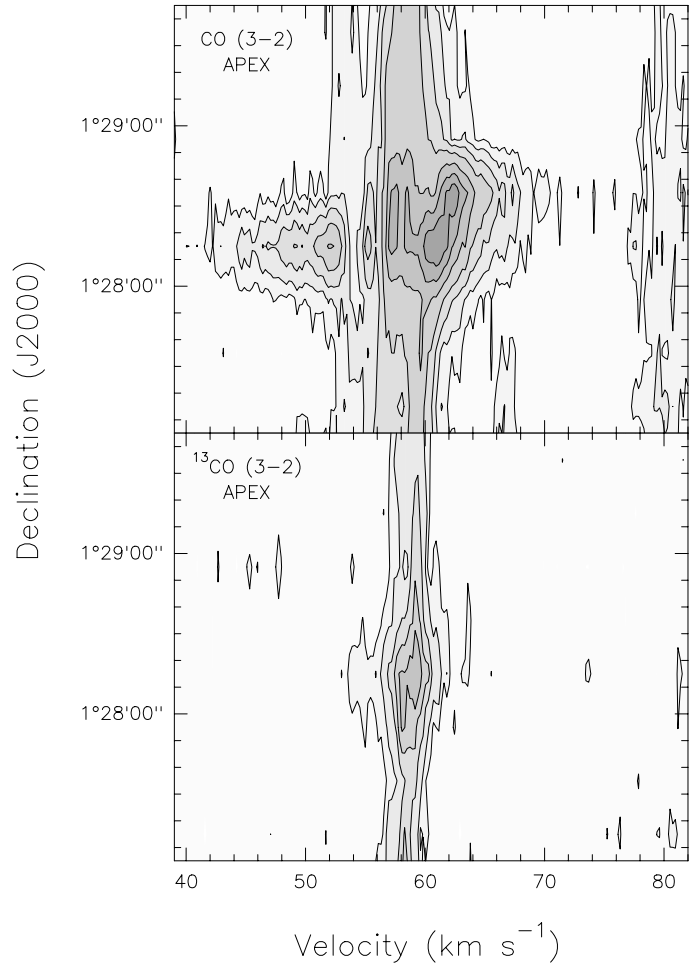}

\end{center}

\caption{Position-velocity diagrams of \dco\ and \tco\ toward the
 northernmost region of G34.43+0.24 along a line running since the
north-south direction passing through
of MM3 (position (40\arcsec,~0\arcsec) in Figure~\ref{fig-13co32-spectra}).
Transitions and
telescopes are indicated in the top left corner of each panel.
The contour levels for \dco\ are 3, 6, 10, 15, 22, 29 and 36 $\sigma$,
 with $\sigma$ = 0.27 K. For \tco\ are 1.5, 3, 6, 8 and 10 $\sigma$, with
$\sigma$ = 0.37 K.}

\label{fig-wingpos-vel_MM3}
\end{figure}

Figure~\ref{fig-wingpos-vel_MM3} shows position-velocity diagrams
 of the \dco\ and \tco\ emission toward the northern region of
 G34.43+0.24, along a line in the north-south direction passing
 through the peak position of the MM3 core.  Clearly seen in the
 \dco\ cut is the presence of wing emission emanating from MM3,
 with velocities extending up to 42.2 \kms\ in the blueshifted side
 and 69.2 \kms\ in the redshifted side. In the \tco\ line the wing
 emission is barely seen. 

\begin{figure}

\begin{center}

\includegraphics[angle=0,scale=0.9]{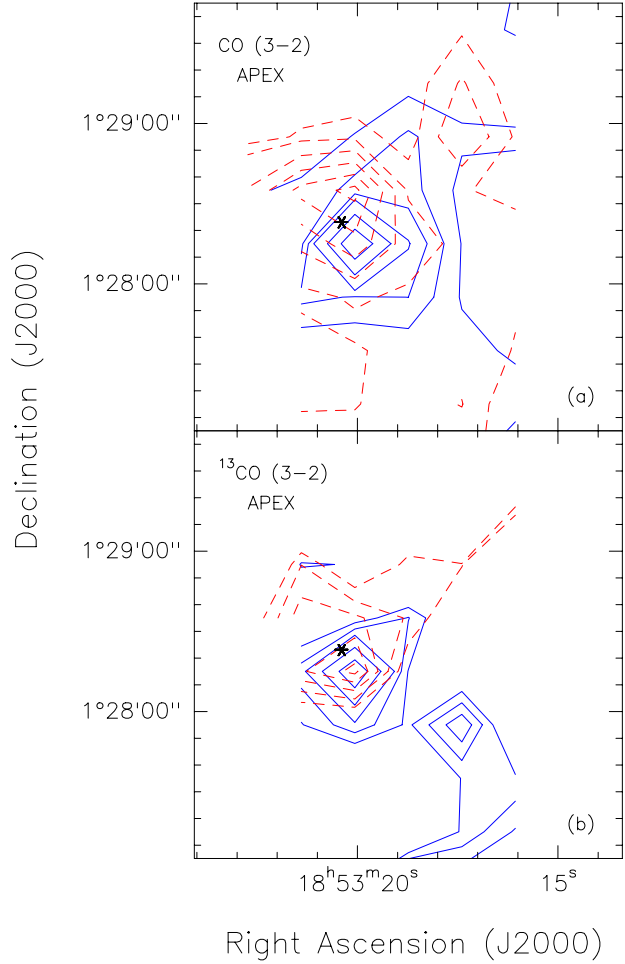}

\end{center}

\caption{Contour maps of velocity-integrated wing emission of \dco\ and
\tco\ toward the northernmost region of G34.43+0.24. Solid lines represent
 blueshifted emission and dashed lines represent redshifted emission. The
 asterisk shows the position of the MM3 core.
{\it Top}: \dco\ emission. Contour levels for the blueshifted emission are
20, 30, 40, 50, 75 and 100 $\sigma$ ($\sigma$ = 0.35 K \kms);
 and for redshifted emission, these 
 are 20, 30, 60, 90, 120 and 150 $\sigma$ ($\sigma$ = 0.25 K \kms). 
{\it Bottom}: \tco\ emission. Contour levels for the blueshifted emission are
3, 4, 5, 7 and 9 $\sigma$ ($\sigma$ = 0.23 K \kms);
 and for redshifted emission, these  
are 3, 4, 5, 6 and 7 $\sigma$ ($\sigma$ = 0.19 K \kms).}
\label{fig-wingpos-pos_MM3}

\end{figure}

Figure~\ref{fig-wingpos-pos_MM3} shows maps of the integrated line
 wing emission, in the \dco\ and \tco\ lines, toward the MM3 core.
 The range of velocity integration of the blueshifted emission (blue
 contours) is $41.2<v_{lsr}<56.1$ \kms\ for \dco\ and
 $52.6<v_{lsr}<56.1$ \kms\ for \tco; whereas that of the redshifted
 emission (red contours) is $61.5< v_{lsr}<69.4$ \kms\ for \dco\
 and $61.5< v_{lsr}<63.8$ \kms\ for \tco. In the \dco\ map, the
 peak positions of the redshifted and blueshifted wing emissions
 are located, respectively, NE and SW of the peak position 
of the millimeter dust core. The red and blue wing emissions exhibit 
however a considerable overlap in their spatial distribution. 

\section{Discussion}

\subsection{Molecular Cores}

\subsubsection{Physical Parameters}

From the observations of the \tco\ and \cdo\ lines it is possible
to estimate the column density and the mass of molecular gas within the 
central region (MM1 and MM2 cores) as follows.  
Assuming that the \cdo\ line is emitted under local 
thermodynamic equilibrium (LTE) conditions and its emission is optically 
thin, then the column density is given by (e.g., Bourke et al. 1997)
\begin{equation}
N({\rm C}^{18}{\rm O}) = 8.084\times 10^{13}~
\frac{(T_{ex}+0.88)\exp(15.81/T_{ex})}{1-\exp(-15.81/T_{ex})}
 \int\!\!\int \tau_{18} dv~,
\label{eqn-lte-column}
\end{equation}
where $\tau_{18}$ is the optical depth of the \cdo\ line, $T_{ex}$ is the 
excitation temperature, and $v$ is in \kms. The LTE mass can be obtained from 
\begin{equation}
\left(\frac{M_{LTE}}{M_{\odot}}\right) = 0.565
\left(\frac{\mu_m}{2.72 m_H}\right)
\left(\frac{[{\rm H}_2/{\rm C}^{18}{\rm O}]}{3.8\times10^{6}}\right)
\left(\frac{D}{\rm kpc}\right)^2 \,
\frac{(T_{ex}+0.88)\exp(15.81/T_{ex})}{1-\exp(-15.81/T_{ex})}
 \int\!\!\int \tau_{18} dv d\Omega~,
\label{eqn-lte-mass}
\end{equation}
where $\mu_m$ is the mean molecular mass per H$_2$ molecule, $m_{\mathrm{H}}$ 
is the mass of a hydrogen atom, $D$ is the source distance, 
[H$_2$/C$^{18}$O] is 
the H$_2$ to C$^{18}$O abundance ratio, and $\Omega$ is in arcmin$^2$. The 
opacities in the \cdo\ line are obtained from the ratio of the observed 
brightness temperatures in the \cdo\ and \tco\ lines as follows (see Bourke 
et al. 1997, for derivation)
\begin{equation}
\frac{1-\exp{(-\tau_{18}/r)}}{1-\exp{(-\tau_{18})}} = \frac{[J_{18}(T_{ex})-J_{18}(T_{\rm bg})]}{[J_{13}(T_{ex})-J_{13}(T_{\rm bg})]}\,\frac{T_{\rm mb}(^{13}{\rm CO})}{T_{\rm mb}({\rm C}^{18}{\rm O})}~,
\label{eqn-for-tau}
\end{equation}
where $T_{\rm bg}$ is the background temperature, $T_{\rm mb}$ is the main beam 
brightness temperature of the line, $J$ is given by (the subscripts  
``18'' and  ``13'' refer to the C$^{18}$O and $^{13}$CO isotopes, respectively)
\begin{equation}
J(T) = \frac{h\nu}{k}\frac{1}{\exp(h\nu/kT)-1}~,
\label{eqn-J}
\end{equation}
and, $r=\tau_{18}/\tau_{13}$ is the optical depth ratio
 \begin{equation}
\frac{\tau_{_{18}}}{\tau_{_{13}}} =
\left[\frac{\rm{C}^{18}\rm{O}}{^{13}\rm{CO}}\right]\frac{(kT_{ex}/hB_{13} +
  1/3)}{(kT_{ex}/hB_{18} + 1/3)}\,\frac{\exp (E_{_{J_{13}}}/kT_{ex})}{\exp
  (E_{_{J_{18}}}/kT_{ex})}\,\frac{[1 - \exp (-h\nu_{_{18}}/kT_{ex})]}{[1 - \exp
  (-h\nu_{_{13}}/kT_{ex})]}~,
\label{eqn-razon-taus}
\end{equation}
where $B$ is the rotational constant of the molecule, $E_{_J}=hBJ(J+1)$ is the 
rotational energy at the lower state, and $J$ is the rotational quantum number. 
Assuming an [$^{13}$CO/C$^{18}$O] abundance ratio of 7.6, we derive average optical 
depths in \cdo\ of 0.10 and 0.18 toward the MM1 and MM2 cores, respectively.
Using an excitation temperature of 30 K and an [H$_2$/C$^{18}$O]
abundance ratio of $3.8\times10^6$ (Wilson \& Rood 1994) we derive total 
masses of 330 and 1460 \Msun\ for the MM1 and MM2 cores, respectively.
The derived core parameters are listed in Table~\ref{tbl-derivparams1}.

Masses can also be estimated assuming that the clouds are in virial
equilibrium. Neglecting magnetic fields and external forces, the virial mass, 
$M_{vir}$, for a spherical cloud of radius $R$ is given by (MacLaren, 
Richardson \& Wolfendale 1988)
\begin{equation}
M_{vir} = B \left({{R}\over{\rm pc}}\right)
\left({{\Delta v}\over{\rm km~s^{-1}}}\right)^2 ~M_{\odot}~,
\label{eqn-virial-mass}
\end{equation}
where $\Delta v$ is the average line width in \kms, $R$ in pc, and $B$ is a
 constant which depends on the density profile of the cloud.
 Assuming that the cores have uniform densities ($B=210$) we derive
 virial masses of $1.1\times10^3$, $1.5\times10^3$ and 
$1.4\times10^3$ \Msun\ for, respectively, the MM1, MM2 and MM3 cores. 
These values correspond to the geometric mean of the masses determined 
from the transitions given in 
Table~\ref{tbl-derivparams2}. The molecular hydrogen densities were computed
 from the mass assuming that the cloud has a spherical morphology and a  
uniform density. 

Table~\ref{tbl-comparison} summarizes the core masses determined  
from different methods and authors. Cols. (2) and (3) give the LTE and 
the virial masses, respectively, determined in this work; col. (4) gives 
the LTE mass from Shepherd et al. (2007); and, cols. (5) and (6) give the 
masses derived from dust continuum observations by Garay et al. (2004)
 and Rathborne et al. (2006), respectively.
In our estimate, if the excitation temperature is 25 or 35 K, the LTE mass 
is decreased or increased by $\sim$20\%, respectively.
 Given the uncertainties 
associated with each method -- such as in the abundance, excitation 
temperature, dust temperature, and dust opacity -- the agreement is 
good between the masses 
determined in this work and the masses determined from dust emission. 
Differences in the mass obtained by different authors using 
dust emission are mostly due to different dust temperatures and core 
sizes used in their calculations. There are, however, 
discrepancies with the masses determined by Shepherd et al. (2004)
from H$^{13}$CO$^+(1\rightarrow0)$ observations, of 4000 and 
5000 \Msun\ for MM1 
and MM2, respectively, and a total cloud mass of 50000 \Msun. 
We note that Rathborne et al. (2005) estimated from dust emission observations  
a total cloud mass of 7500 \Msun\ assuming a dust temperature of 30 K. 
A possible explanation of the larger mass estimate of Shepherd et al. (2004) 
is that the actual abundance of HCO$^+$ is probably larger than the adopted
one, due to an enhancement produced by the molecular outflows.  
Moreover, Shepherd et al. (2007) determined, from observations of the 
C$^{18}$O$(1\rightarrow0)$ line using the OVRO array, masses of 75 and 
690 \Msun\ for MM1 and MM2, respectively. These values probably correspond to 
the inner mass of the cores. We note that masses derived from molecular 
lines might not be tracing all of the emission, and their emission may be 
optically thick, specially C$^{18}$O$(1\rightarrow0)$. However, the 
assumption that the \cdo\ is optically thin seems to be adequate and it is 
supported for the inspection of the line profiles and values obtained for 
the optical depths. 

Optical depths and column densities toward the peak position of
the MM1 and MM2 cores are given in Table~\ref{tbl-derivparampeak}.
The \cstyc\ and \csdos\ values were obtained in an analogous way to that
described above for the \cdo\ and \tco\ lines, adopting a [CS/C$^{34}$S]
abundance ratio of 22.5 and an [H$_2$/CS] ratio of $1\times10^8$ (e.g., van
der Tak et al. 2000).

\subsubsection{MM2: a Collapsing Core}
\label{infall}

The \tco\ line emission at the peak position of the MM2 core shows a bright 
blueshifted peak and a flat redshifted shoulder. Such line shape 
can be modeled as due to inward motions (Mardones et al. 1997). 
The blue peak is produced by high excitation gas on the far 
side of the envelope, while the redshifted peak results from low 
excitation gas in the near side of the envelope. The self--absorption due 
to low excitation, low velocity gas in the near side of the envelope 
gives rise to the central dip (Evans 1999). An estimate of the infall 
velocity, $V_{in}$, can be obtained using the two--layer infall model 
of Myers et al. (1996).
 
\begin{figure}
\begin{center} 
\includegraphics[angle=0,scale=0.9]{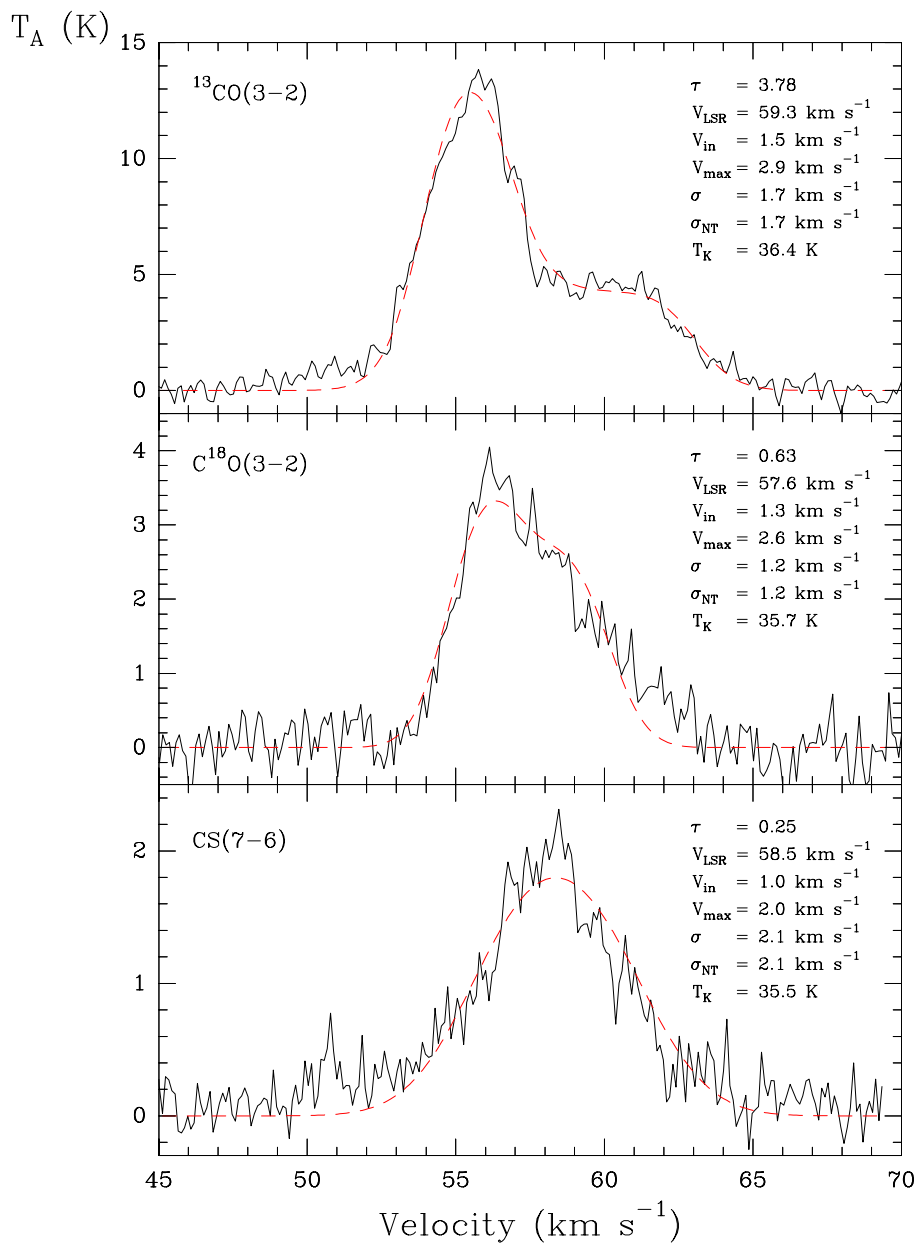}
\end{center}
\caption{\tco, \cdo, and \cssiete\ spectra of the MM2 peak position with the 
simple two layer model fits of Myers et al. (1996). The solid lines show the 
observed spectra and the dashed (red) lines represent the model. The values
derived by the model are shown in the right side of the figure.}
\label{fig-Myers-model}
\end{figure}

Myers et al. (1996) made a simple analytic model of radiative transfer in 
order to inspect if the observed spectra show any evidence of infall 
motions and to provide an estimate of the characteristic $V_{in}$.
They assumed two uniform regions of equal temperature and velocity 
dispersion $\sigma$, whose density and velocity are attenuation-weighted 
means over the front and rear halves of a centrally condensed, 
contracting cloud with 
$n/n_{max}=(r/r_{min})^{-3/2}$ and $V/V_{max}=(r/r_{min})^{-1/2}$. We modeled 
the profile of the \tco, \cdo\ and \cssiete\ lines toward the MM2 core 
using their prescription and obtained the following model parameters: the peak 
optical depth, $\tau_0$; the kinetic temperature, $T_k$; the infall velocity, 
$V_{in}$; and the non--thermal velocity dispersion, $\sigma_{_{NT}}$. 
Figure~\ref{fig-Myers-model} shows the observed and the best model 
line profiles, indicating motions consistent with a ``fast'' infall, 
$V_{in}\sim \sigma$. The values derived by the model are shown in the top right
corner in the same figure. We note that some of the parameters for each fit
are different, for example, the infall velocities are 1.5, 1.3, and 1.0
\kms\ for $^{13}$CO, C$^{18}$O, and CS, respectively. This decreasing order is
in accord with the level of asymmetry in the lines. The most asymmetric
line, a single peak in the blue and a red shoulder, has the largest values of 
$V_{in}$ (Myers et al. 1996; Fuller et al. 2005). Although the velocity 
dispersion in the CS line 
is $\sigma \sim 2V_{in} $, the other two lines have $\sigma \sim V_{in}$
indicating a fast infall. The derived systemic velocities show some scatter,
 most likely due to the influence of the molecular outflow, which 
is more intense toward redshifted velocities. If the contribution of
the outflow is removed, the blue peak becomes stronger than the 
red shoulder, and this will have the effect of increasing the derived infall
velocity.  

Ramesh et al. (1997) modeled the profile of the \csdos, \cstyc, \hco\ and 
\hcotrece\ lines toward the MM2 core using both the large velocity
 gradient (LVG) 
method and the simple model of Myers et al. (1996). For the LVG method they 
adopted a constant temperature of 22 K (we determined a kinetic temperature of 
$\sim$36 K), and found a collapse velocity that varies radially as 
$8 (r/r_{0})^{-0.6}$ \kms\ with $r_{0}=4.5\times10^{16}$ cm. At a radius of 
$r=5.5\times10^{17}$ cm (equivalent to an angular size of 10\arcsec) we get 
$V_{in}=1.8$ \kms. Given the uncertainties associated with each method, the 
agreement is good with the values determined by us. The infall speed 
estimated by Ramesh et al. (1997) using the Myers et al. (1996) analytic
model is about 1 order of magnitude smaller than the velocities they  
determine from the LVG analysis; however,  since the fit of the Myers 
model is admittedly not too good, they argue in favor of the larger value,
 which agrees well with ours here.

The mass infall rate can be estimated assuming $\dot M = M_R V_{in}/R$, 
where $M_R$ is the mass within the radius $R$ and $V_{in}$ is the infall
 speed at
radius $R$. Using the LTE mass derived at the central position 
of MM2, of 240 \Msun, an average infall velocity of 
1.3 \kms, and a radius of $5.5\times10^{17}$ cm (corresponding 
to a 10\arcsec angular radius); we obtain a mass infall rate 
of $1.8\times10^{-3}$ \Msun\ yr$^{-1}$. Such accretion rate is high enough to
overcome the radiation pressure of the central star and to form a massive 
star. Shepherd et al. (2007) identified two likely massive
stars (M $>$ 8 \Msun) with the MM2 core, with one of them, located near the
projected center of the UC H II region, having a spectral type of B0.5 
(Shepherd et al. (2004).  Models of massive envelopes accreting onto this kind 
of young stars require accretion rates of
 $\dot M > 5\times10^{-4}$ \Msun\ yr$^{-1}$ (Osorio, Lizano \& D'Alessio 
1999), in agreement with our results.    

\subsection{Molecular Outflows}

To compute physical parameters of the outflowing gas, we followed the standard
LTE formalism described in Bourke et al. (1997) using the emission in the
CO($3\rightarrow2$) line. The main source of error in determining the mass
 in the outflow arises from the difficulty in separating the contribution
 to the outflows of the emission in the velocity range of the ambient cloud. 
We adopt as velocity boundary between the blue and red wing emission
and the ambient emission the values of, respectively, $53.0$ and $60.8$ \kms\
for the MM1 outflow, $51.8$ and $60.9$ \kms\ for the MM2 outflow, and
56.1 and 61.5 \kms\ for the MM3 outflow. Another source of error is the 
possibility that the \dco\ line might not be optically thin, particularly
 at low flow velocities.  Fortunately, with the available data, we can
 asses this effect and make corrections.

The mass in the high velocity wings is estimated from the CO(3$\rightarrow$2)
observations as follows. Assuming that the energy levels of CO are populated
according to LTE and that the outflow emission is
optically thin, then the total CO column density, $N({\rm CO})$, at each
observed position is given by (e.g., Garden et al. 1991)
\begin{equation}
  N({\rm CO})=7.7 \x 10^{13}\,\, \frac{(T_{ex}+0.92)\,\exp(16.60/T_{ex})}{[1-\exp(-16.60/T_{ex})]} \,\,\frac{1}{[J(T_{ex}) - J(T_{bg})]} \int T_{B}(v)dv~,
\end{equation}
where $T_{ex}$ is the excitation temperature of the population
in the rotational levels, $T_{bg}$ is the background
temperature, $T_{B}$ is the brightness temperature of the CO(3$\rightarrow$2)
emission at velocity $v$, $v$ is measured in \kms, and $J$ is defined by 
equation~(\ref{eqn-J}).
The mass is then computed from the derived column densities as,
\begin{equation}
M = [{\rm H}_2/{\rm CO}]~\mu_m~A~ \sum N({\rm CO})~,
\end{equation}
where $\mu_m$ is the mean molecular mass per H$_2$ molecule, [H$_2$/CO] is 
the molecular hydrogen to carbon monoxide abundance ratio, $A$ is the 
size of the emitting area in an individual position, and the sum is 
over all the observed positions inside each lobe. 

From the observations of the \dco\ and \tco\ lines it is possible to estimate 
the opacities of the flowing gas and therefore to assess whether or not the 
assumption that the \dco\ wing emission is optically thin is correct. Using 
expressions (\ref{eqn-for-tau}) and (\ref{eqn-razon-taus}) for $^{12}$CO 
and $^{13}$CO, assuming $T_{ex}$ = 30 K, and [$^{12}$CO/$^{13}$CO] = 50, 
we find that the average \dco\ opacities of the 
blueshifted gas (in the velocity range from 49.5 to 53.0 \kms) and 
redshifted gas (in the velocity range from 60.8 to 68.9 \kms) of the 
MM1 flow are 3.4 and 2.5, respectively. 
The given ranges are those in which emission is detected in both lines. For the 
MM2 outflow we find that the average optical depth of the blueshifted gas 
(in the velocity range from 47.2 to 51.8 \kms) and redshifted gas (in the
 velocity range from 60.9 to 68.7 \kms) are 5.4 and 4.4, respectively. For the
 MM3 outflow we find that the average optical depth of the blueshifted gas
 (velocity range from 52.8 to 56.1 \kms) and redshifted gas (velocity range
 from 61.5 to 65.2 \kms) are 7.1 and 4.2, respectively. These calculations
 indicate that the \dco\ emission from the flowing gas, in the given velocity
 ranges, is moderately optically thick for the three flows. On the other 
hand, as the position--velocity diagrams suggested, the $^{13}$CO flow 
emission is optically thin, with optical depths between 0.05 and 0.13.  

To correct the mass determined under the optically thin assumption by the
opacity effect we divided the \dco\ wing emission into two regimes, an
optically thin regime, in which no \tco\ emission is detected, and an
optically thick regime. We assumed an [H$_2$/CO] abundance ratio of 
$10^4$ (Frerking, Langer \& Wilson 1982) and an excitation temperature of the 
outflowing gas of 30 K.  We derived masses in the optically thin regime of the  
0.5, 1.1, and 1.6 \Msun\ for the blueshifted lobes of 
MM1 ($38.6<v_{\rm lsr}<49.5$ \kms), MM2 ($35.9<v_{\rm lsr}<47.2$ \kms), 
and MM3 ($40.1<v_{\rm lsr}<52.8$ \kms), respectively, 
and masses of 0.7, 0.8, and 0.7 \Msun\ for the redshifted lobes of  
MM1 ($68.9<v_{\rm lsr}<74.5$ \kms), MM2 ($68.7<v_{\rm lsr}<74.5$ \kms), 
and MM3 ($65.2<v_{\rm lsr}<74.3$ \kms), respectively. 
In the optically thick regime the mass determined under the optically 
thin assumption is multiplied by the 
factor $\tau_{\mathrm{w}}/(1-e^{-\tau_{\mathrm{w}}})$, where $\tau_{\mathrm{w}}$
is the average \dco\ optical depth of wing emission in the optically thick 
velocity range. We derive masses in the optically thick regime,
corrected by opacity, of 2.6, 13.6, and 8.5 \Msun\ for the blueshifted lobes 
of MM1 ($49.5<v_{\rm lsr}<53.0$ \kms), MM2 ($47.2<v_{\rm lsr}<51.8$ \kms), 
and MM3 ($52.8<v_{\rm lsr}<56.1$ \kms), respectively. Similarly, for the 
redshifted lobes we derived masses of 11.9, 30.8, and 7.3 \Msun\ toward MM1 
($60.8<v_{\rm lsr}<68.9$ \kms), MM2 ($60.9<v_{\rm lsr}<68.7$ \kms), and 
MM3 ($61.5<v_{\rm lsr}<65.2$ \kms), respectively.

To estimate the contribution to the flow mass from the low velocity material
emitting in the same velocity range as the ambient cloud gas, we followed
the prescription of Margulis \& Lada (1985).
Using expression (A16) of Bourke et al. (1997),
\begin{equation}   
\int_{{\rm lobe}} \int_{v_b}^{v_r} T_{{\rm mb}}\, dv\, d\Omega = \int_{{\rm lobe}}\left(\frac{T^b + T^r}{2}\right)(v_r - v_b)\,d\Omega~,
\label{eqn-mass-hidden}
\end{equation}
where $T^b$, $v_b$ and $T^r$, $v_r$ are the brightness temperature and velocity 
at the blue and red velocity boundaries, respectively.  We estimated that 
the hidden flow masses corrected by opacity in the low
velocity range within the blueshifted lobes toward MM1, MM2 and MM3 are
3.0, 20.6 and 13.9 \Msun, and within the redshifted lobes are 6.2, 11.9
and 8.9 \Msun, respectively (see Table~\ref{tbl-derivparams-outflows}). 
Finally, the total mass of the outflows are: 24.9 \Msun\ for MM1, 78.8 \Msun\ 
for MM2, and 40.9 \Msun\ for MM3.

Using the standard LTE formalism, it is also possible to estimate the momentum,
$P$, and the kinetic energy, $E_{\rm{k}}$, in the 
flow as follows (Margulis \& Lada 1985; Bourke et al. 1997)
\begin{eqnarray}    
 P & = & \int \!\! \int \!\! \int M(v,\alpha,\delta)\,v\,dv\,d \alpha \,d\delta~, \nonumber\\
E_k & = & \frac{1}{2}\int \!\! \int \!\! \int M(v,\alpha,\delta)\,v^2\,dv\,d \alpha \,d\delta~.
\label{eqn-LTE-P_Ek}
\end{eqnarray}

 These equations assume no
correction for flow inclination, and so are strict lower limits. Another method
to compute the physical parameters of the outflowing gas is to assume that all
the mass is flowing at a velocity characteristic of the entire flow, 
$V_{\rm{char}}$. The flow parameters are, in this approach, given by
\begin{equation}
P = M~V_{\mathrm{char}}~,~~~~
E_{\mathrm{k}} = \frac{1}{2}~M~V_{\mathrm{char}}^2~,  
\end{equation}
where $M$ is the total mass of the outflow lobe. 
The blueshifted and redshifted emissions from the outflows show considerable 
overlap in their spatial distribution, indicating that the symmetry axis of the 
flows is likely to be near the line of sight. Hence, we adopt as a
 characteristic 
flow velocity the average of the maximum observed outflow velocities. The 
parameters determined under this assumption are likely to correspond to upper 
limits. The maximum velocities observed toward the blue and red lobes
are, respectively, 12.0 and 23.1 \kms\ for the MM1 outflow, 15.0 and 15.9 \kms\
for the MM2 outflow, and 17.3 and 11.6 \kms\ for the MM3 outflow.
Table~\ref{tbl-derivparams-outflows} gives the derived parameters of the 
blueshifted gas and redshifted gas of the flows calculated using both methods.
The geometric mean values for the outflow parameters are 
$\bar {P_1} \sim 230$, $\bar P_2 \sim 690$, and $\bar P_3 \sim 270$ \Msun\ 
\kms; and $\bar E_{{\rm k}_1} \sim 1290$, $\bar E_{{\rm k}_2} \sim 3630$, and
$\bar E_{{\rm k}_3} \sim 1120$ \Msun\ km$^2$ s$^{-2}$ 
(subscripts 1, 2, and 3 refer to the MM cores). If the excitation temperature 
is increased to 50 K, the flow parameters increase by 7\%. 
The derived values of the outflow parameters are similar to those of massive and 
energetic molecular outflows driven by high--mass young stellar objects (Shepherd 
\& Churchwell 1996; Beuther et al. 2002; Wu et al. 2004).   
We do not provide estimates of the dynamical 
timescale, mechanical luminosity, and mass outflow rate of the outflows 
toward G34.43+0.24 because they are likely to be aligned close to
 the line of sight, and therefore the lengths of the lobes are difficult
 to estimate. 

Rathborne et al. (2005) reported an excess 4.5 $\mu$m emission toward
the MM cores. The {\it Spitzer} IRAC 4.5 $\mu$m band contains the H I
Br$\alpha$ line (at 4.052 $\mu$m), the CO$(1\rightarrow0)$ P(8) band 
head (4.6--4.8$\mu$m), and many H$_2$ vibrational transitions 
(the most important being H$_2(0\rightarrow0)$ S(9) line at 4.694 $\mu$m), 
but unlike the other three bands it does not 
contain any strong polycyclic aromatic hydrocarbon (PAH) features 
(e.g., Smith et al. 2006). It has been postulated that 
the main contributers of the excess 4.5 $\mu$m emission could be 
emission from the CO$(1\rightarrow0)$ P(8) band head (e.g., Marston et al. 
2004) and/or emission from the H$_2(0\rightarrow0)$ S(9) line (e.g., 
Noriega--Crespo et al. 2004). Thus, the enhancement of the 4.5 $\mu$m emission 
could be a probe of jets and molecular outflows (e.g., Teixeira et al. 2008; 
Smith et al. 2006; Noriega--Crespo et al. 2004); therefore of current star 
formation. Rathborne et al. (2005) first found evidence for the presence
of high velocity gas toward the MM1 and MM3 cores from the broad line 
widths ($\Delta V \sim$10 \kms) observed in the CS$(3\rightarrow2)$ and 
HCN$(3\rightarrow2)$ lines. A detailed 
study of the central region of G34.43+0.24 was carried out by Shepherd et al. 
(2007), who reported two molecular outflows emanating from MM1 
(G34.4 MM, in their paper), and three outflows originating from 
the MM2 core. They estimate a total mass, 
momentum and kinetic energy of 34.8 \Msun, 350 \Msun\ \kms, and 
$4.9\times 10^{46}$ ergs (2460 \Msun\ km$^2$ s$^{-2}$), for the outflows 
in MM1; and 111.2 \Msun\ , 830 \Msun\ \kms, and $7.8\times 10^{46}$ ergs 
(3920 \Msun\ km$^2$ s$^{-2}$), for the outflows in MM2. The masses are 
$\sim$20\% greater than our estimate, and the momenta and kinetic
 energies are within our estimated range. In spite of the different molecular 
transitions, angular resolutions and assumptions used, 
Shepherd et al. and our results are in agreement, within the 
observational errors.

\subsubsection{The Discovery of a New Molecular Outflow Toward the Core 
G34.43+0.24 MM3}

We report the discovery of one massive molecular outflow associated with 
the MM3 core discovered by Garay et al. (2004). They suggested
that this massive and dense core, without counterpart at mid--infrared 
({\it MSX}) and far--infrared ({\it IRAS}) wavelengths implying that it has
a low temperature, seemed to be in a quiescent state, 
and thus appeared to be the youngest of the massive cores in the filament. 
The presence of 24 $\mu$m emission, a green fuzzy, water and methanol masers 
(Rathborne et al. 2005; Chambers et al. 2009) and  
the strong wing emission in the \dco\ line, indicating the presence of a 
powerful molecular outflow, all indicate star formation activity in the
northernmost region of the filament. Due to its spatial location and 
mass, YSO number 29 in Shepherd et al. (2007) appears to be a
 good candidate for being the driver of the MM3--outflow. It is
 the most massive stellar object within MM3, with a mass of $\sim$4.8--6.8 
\Msun\ and a luminosity of 
$\sim$170--340 \Lsun. However, this is an intermediate mass star that 
may not be able to drive alone the massive outflow. In the same region
there are three more stars of intermediate mass (less massive than the 
number 29) with maximum masses between 4--5 \Msun. 
Interferometric observations are needed to clarify if the MM3--outflow 
corresponds to an overlap of a few individual outflows, such as that seen in 
the central region of G34.43+0.24, or it is driven by a single star.   

\section{Conclusions}

We carried out a multi--line study of cores within the filamentary 
IRDC  G34.43+0.24. The emission from the filament was 
fully mapped in the \dco\ and 
\tco\ lines. In addition, emission from the central region of 
the filament was observed in eight different molecular transitions 
[\cdo, \cssiete, \csdos, \cstyc, \hco, \hcotrece, \sio\ and \metanol], and 
from the north region in the \csdos\ and \cdo\ lines. The results of this
study are summarized as follows. 

The spatial distribution of the molecular emission in all, except methanol, 
lines is similar to that of the dust continuum emission derived 
from observations with similar angular resolution, showing a filamentary 
structure and four cores. 

The mass of the molecular cores derived assuming LTE conditions 
are 300 and 1460 \Msun\ for the MM1 and MM2 cores, respectively.
The masses derived assuming virial equilibrium are 1100, 1500 and
 1400 \Msun\ for the MM1, MM2 and MM3 cores, respectively.
The average molecular hydrogen densities of the cores 
are $1.8\times 10^5$ cm$^{-3}$ in MM1, $1.7\times 10^5$ cm$^{-3}$ 
in MM2 and $1.9\times10^4$ cm$^{-3}$ in MM3. We also find that the 
molecular hydrogen column densities at the peak positions of MM1 
and MM2 are, respectively, $2\times 10^{22}$ cm$^{-2}$ and $8\times 10^{22}$ 
cm$^{-2}$.  The derived parameters of the MM1 and MM2 cores are typical 
of massive and dense cores harboring high-mass YSO's (Plume et al. 1997;
 Garay et al. 2007). Those of the MM3 core suggest that it harbors young 
intermediate mass stars, in accord with the result obtained by 
Shepherd et al. (2007) from the modeling of the SED of the NIR 
sources detected using {\it Spitzer}.   
 
We modeled the molecular emission in the \tco, \cdo, and \cssiete\ lines 
toward the MM2 core, concluding that this core is undergoing infalling 
motions, with an infall velocity of $\sim$1.3 \kms\ and 
a mass infall rate of $\sim$$1.8\times 10^{-3}$ \Msun\ yr$^{-1}$. This value 
is large enough to allow the formation of massive stars by accretion
(see Osorio, Lizano \& D'Alessio 1999). 

We report the discovery of a molecular outflow associated with the massive, 
dense MM3 core located in the northern region of G34.43+0.24, and 
possibly of a second one associated with the MM4 core. 
The molecular outflow within MM3 has a total mass of 41 \Msun, a momentum of 
270 \Msun\ \kms, and a kinetic energy of 1120 \Msun\ km$^2$ s$^{-2}$; 
indicating that it is driven by high--intermadiate mass stars. We also 
detected high velocity molecular gas toward the MM1 and MM2 cores with total 
masses of 25 and 80 \Msun, momenta of 230 and 690 \Msun\ \kms, and kinetic 
energies of 1290 and 3630 \Msun\ km$^2$ s$^{-2}$, respectively. 
     
\acknowledgements

The authors gratefully acknowledge support from CONICYT projects  
FONDAP No. 15010003 and BASAL PFB-06.

\begin{center}
{\bf REFERENCES}
\end{center}

Beuther, H., Schilke, P., Sridharan,
  T. K., Menten, K. M., Walmsley, C. M., \& Wyrowski, F. 2002, \aap, 383, 892 

Bourke, T. L., Garay, G.,
Lehtinen, K. K., K\"ohnenkamp, I., Launhardt, R., Nyman, L.-\AA., May, J.,
Robinson, G., \& Hyland, A. R. 1997, \apj, 476, 781

 Bronfman, L., Nyman, L.-\AA., \& May, J. 1996, \aaps, 115, 81

 Carey, S.~J., Clark, 
F.~O., Egan, M.~P., Price, S.~D., Shipman, R.~F., 
\& Kuchar, T.~A.\ 1998, \apj, 508, 721 

 Carey, S.~J., Feldman, 
P.~A., Redman, R.~O., Egan, M.~P., MacLeod, J.~M., 
\& Price, S.~D.\ 2000, \apjl, 543, L157
 
 Chambers, E.~T., 
Jackson, J.~M., Rathborne, J.~M., \& Simon, R.\ 2009, \apjs, 181, 360 

 Cortes, P., Crutcher, R. M., Shepherd, 
D. S., \& Bronfman, L. 2008, \apj, 676, 464 

 Egan, M.~P., Shipman, R.~F., 
Price, S.~D., Carey, S.~J., Clark, F.~O., \& Cohen, M.\ 1998, \apjl, 494, L199
 
 Evans, II, N. J. 1999, \araa, 37, 311 

 Fa\'undez, S., Bronfman, L.,
Garay, G., Chini, R., Nyman, L.-\AA., \& May, J. 2004, \aap, 426, 97

 Frerking, M. A., Langer, W. D., \&
  Wilson, R. W. 1982, \apj, 262, 590

 Fuller, G. A., Williams, S. J., \&
  Sridharan, T. K. 2005, \aap, 442, 949

 Garay, G., \& Lizano, S.\ 1999, 
\pasp, 111, 1049 

 Garay, G., Fa\'undez, S., Mardones, D.,
Bronfman, L., Chini, R., \& Nyman, L.-\AA. 2004, \apj, 610, 313

 Garay, G., Mardones, D., Brooks, K. J.,
Videla, L., \& Contreras, Y. 2007, \apj, 666, 309

 Garden, R. P., Hayashi, M., Hasegawa,
  T., Gatley, I., \& Kaifu, N. 1991, \apj, 374, 540

 Hartquist, T. W., Dalgarno, A.,
  \& Oppenheimer, M. 1980, \apj, 236, 182

 Hennebelle, P., P{\'e}rault, M., 
Teyssier, D., \& Ganesh, S.\ 2001, \aap, 365, 598 

 Jackson, J.~M., Finn, 
S.~C., Rathborne, J.~M., Chambers, E.~T., \& Simon, R.\ 2008, \apj, 680, 349 

 MacLaren, I., Richardson, K. M., \& Wolfendale, A. W. 1988, \apj, 333, 821

 Mardones, D., Myers, P. C.,
  Tafalla, M., Wilner, D. J., Bachiller, R., \& Garay, G. 1997, \apj, 489, 719

 Margulis, M. \& Lada, C. J. 1985, \apj, 299, 925 

 Marston, A.~P., et al.\ 2004, \apjs, 154, 333 

 Miralles, M. P., Rodr\'iguez, L. F., \& Scalise, E. 1994, \apjs, 92, 173

 Molinari, S., Brand, J., Cesaroni, R., \& Palla, F. 1996, \aap, 308, 573

 Molinari, S., Brand, J., Cesaroni,
  R., Palla, F., \& Palumbo, G. G. C. 1998, \aap, 336, 339

 Myers, P. C., Mardones, D.,
  Tafalla, M., Williams, J. P., \& Wilner, D. J. 1996, \apj, 465, L133

 Noriega-Crespo, A.,
  Morris, P., Marleau, F. R., Carey, S., Boogert, A., van Dishoeck, E., Evans,
  II, N. J., Keene, J., Muzerolle, J., Stapelfeldt, K., Pontoppidan, K.,
  Lowrance, P., Allen, L., \& Bourke, T. L. 2004, \apjs, 154, 352 

 Osorio, M., Lizano, S., \& D'Alessio,  P. 1999, \apj, 525, 808 

 Perault, M., et al.\ 1996, \aap, 315, L165 

 Pillai, T., Wyrowski, F., Carey, 
S.~J., \& Menten, K.~M.\ 2006, \aap, 450, 569 

 Plume, R., Jaffe, D. T., Evans, N. J. II,
 Mart\'in--Pintado, J., \& G\'omez--Gonz\'ales, J. 1997, \apj, 476, 730

Ramesh, B., Bronfman,
 L., \& Deguchi, S. 1997, \pasj, 49, 307

 Rathborne, J. M., Jackson, J. M.,
 Chambers, E. T., Simon, R., Shipman, R., \& Frieswijk, W. 2005, \apj,
 630, L181

 Rathborne, J. M., Jackson, J. M., \and Simon, R. 2006, \apj, 641, 389

 Rathborne, J.~M., Simon, R., 
\& Jackson, J.~M.\ 2007, \apj, 662, 1082 

 Rathborne, J.~M., 
Jackson, J.~M., Zhang, Q., \& Simon, R.\ 2008, \apj, 689, 1141 

 Shepherd, D. S., \& Churchwell, E. 1996, \apj, 472, 225 

 Shepherd, D. S., N\"urnberger,
 D. E. A., \and Bronfman, L. 2004, \apj, 602, 850

 Shepherd, D. S., Povich, M. S.,
  Whitney, B. A., Robitaille, T. P., N\"urnberger, D. E. A., Bronfman, L.,
  Stark, D. P., Indebetouw, R, Meade, M. R., \& Babler, B. L. 2007, \apj, 669,
  464 

 Simon, R., Jackson, 
J.~M., Rathborne, J.~M., \& Chambers, E.~T.\ 2006a, \apj, 639, 227

 Simon, R., Rathborne, J.~M., 
Shah, R.~Y., Jackson, J.~M., \& Chambers, E.~T.\ 2006b, \apj, 653, 1325  

 Smith, H. A., Hora, J. L., Marengo, M., \& Pipher, J. L. 2006, \apj, 645, 1264 

 Szymczak, M., Hrynek, G., \& Kus, A. J. 2000, \aaps, 143, 269

 Teixeira, P. S., McCoey, C., Fich, M., \& Lada, C. J. 2008, \mnras, 384, 71 

 Turner, J. L. \& Dalgarno, A. 1977, \apj, 213, 386 

 van der Tak, F. F. S., van
  Dishoeck, E. F., Evans, II, N, J., \& Blake, G. A. 2000, \apj, 537, 283

 Wang, Y., Zhang, Q., Rathborne, J. M.,
Jackson, J., \& Wu, Y. 2006, \apj, 651, L125

 Wilson, T. L., \& Rood, R. 1994, \araa, 32, 191 

 Wu,  Y., Wei, Y., Zhao, M., Shi, Y., Yu, W.,
 Qin, S., \& Huang, M. 2004, \aap, 426, 503 

\clearpage

\clearpage

\clearpage

\clearpage

\clearpage

\clearpage

\clearpage

\clearpage

\clearpage

\clearpage

\clearpage

\clearpage

\clearpage

\begin{table}
\begin{center}
\caption{Properties of 1.2 mm Cores \label{tbl-1.2mm}}
\begin{tabular}{cccccc}
\tableline\tableline
       &  \multicolumn{2}{c}{\underline {~~~~Peak Position~~~~}}  &
       \multicolumn{2}{c}{\underline {~FWHM Diameter~}} &    \\
       & $\alpha$(J2000) & $\delta$(J2000) & Angular  & Physical & Mass\\
 Core  &     &     &  (\arcsec) &   (pc) & \Msun \\                          
 \tableline
MM1 & 18 53 18.0 &  01 25 24 & 16 & 0.19 & 1187\\
MM2 & 18 53 18.6 &  01 24 40 & 26 & 0.42 & 1284\\
MM3 & 18 53 20.4 &  01 28 23 & 24 & 0.38 & 301\\
MM4 & 18 53 19.0 &  01 24 08 & 24 & 0.38 & 253\\
\tableline
\end{tabular}
\tablecomments{Units of right ascension are hours, minutes and seconds,
and units of declination are degrees, arcminutes and arcseconds.}
\tablecomments{Properties adopted from Rathborne, Jackson \& Simon (2006).}
\end{center}
\end{table}

\clearpage

\begin{table}
\begin{center}
\scriptsize
\caption{Summary of Observational Parameters \label{tbl-obsparam}}
\begin{tabular}{lccccccc}
\tableline\tableline
                  & Frequency       & Beam (FWHM) &  Spacing  & Observed &Region Size 
 &  $\Delta v$    & rms Noise (T$_{\rm A}$)\\
Observation       & (GHz)           & (arcsec)    &  (arcsec) & Positions&(arcsec$^2$)
 &  (km s$^{-1}$) & (K)       \\
\tableline
\multicolumn{8}{c}{APEX}\\
\tableline
\tco \dotfill     &  330.588        &  18.4       & 20 &
145\tablenotemark{a} & $120\times 700$

 & 0.11           & 0.37      \\
\dco \dotfill     &  345.796        &  17.6       & 20 &
145\tablenotemark{a} & $120 \times 700 $

 & 0.11           & 0.27      \\
\cssiete \dotfill &  342.883        &  17.8       & 20 &
21\tablenotemark{b} & $60 \times 140 $

 & 0.11           & 0.18      \\
\cdo \dotfill     &  329.331        &  18.5       & 20 &
30\tablenotemark{b} & $100 \times 120 $

 & 0.11           & 0.29      \\
\tableline
\multicolumn{8}{c}{Nobeyama}\\
\tableline
\csdos \dotfill   &  97.981         &  15.5      & 20  &
30\tablenotemark{b} & $100 \times 160 $

   & 0.11           & 0.08      \\
\sio \dotfill     &  86.846         &  17.7      & 20  &
30\tablenotemark{b} & $100 \times 160 $

   & 0.13           & 0.07      \\
\cstyc \dotfill   &  96.412         &  15.8      & 20  &
31\tablenotemark{b} & $100 \times 160 $

   & 0.12           & 0.10      \\
\hco \dotfill      &  89.188         &  17.3      & 20  &
31\tablenotemark{b} & $100 \times 160 $

   & 0.12           & 0.08      \\
\hcotrece \dotfill&  86.754         &  17.7      & 20  &
30\tablenotemark{b} & $100 \times 160 $

   & 0.13           & 0.07      \\
\metanol \dotfill &  96.744         &  15.7      & 20  &
31\tablenotemark{b} & $100 \times 160 $

   & 0.12           & 0.09      \\
\tableline
\multicolumn{8}{c}{SEST}\\
\tableline
\csdos \dotfill   &  97.981         &  51.0      & 30  &
25\tablenotemark{c} & $100 \times 100 $

   & 0.13           & 0.07      \\
\cdodos \dotfill  &  219.560        &  24.1      & 30  &
25\tablenotemark{c} & $100 \times 100 $

   & 0.12         & 0.18      \\
\tableline
\end{tabular}

\tablenotetext{a}{Covering the whole filamentary structure.}
\tablenotetext{b}{Covering the central region of G34.43+0.24,
 including the MM1, MM2 and MM4 cores.}

\tablenotetext{c}{Covering the MM3 core, in the northern region 
of the G34.43+0.24 filament.}

\end{center}
\end{table}

\clearpage

\begin{deluxetable}{clccccc}
\tabletypesize{\footnotesize}
\tablecaption{Observed Parameters of Molecular Cores \label{tbl-sizes}}
\tablewidth{0pt}
\tablehead{
\colhead{} &  \colhead{Molecular}  & \colhead{Angular Size}
 & \colhead{$v_{lsr}$} & \colhead{$\Delta V$}  & \colhead{$\int T_{mb} dv$} 
& \colhead{Physical Size}\\
\colhead{Core} &  \colhead{Transition} & ($\arcsec\times\arcsec$)&(\kms)&(\kms)&(K \kms)&(pc$^2$)\\
}
\startdata
\\[-1.1cm]
MM1 & \cssiete \dotfill & $37.2 \times 20.6$  & 57.38 $\pm$ 0.21  &
 5.69 $\pm$ 0.71   & 6.94 $\pm$ 0.80 & $0.66 \times 0.38$\\
    & \csdos \dotfill   & $45.4 \times 33.0$  & 58.13 $\pm$ 0.01  &
 5.29 $\pm$ 0.02    & 15.10 $\pm$ 0.04 & $0.82 \times 0.60$\\
    & \hcotrece \dotfill& $33.0 \times 24.8$  & 58.36 $\pm$ 0.02  &
 2.75 $\pm$ 0.06    & 2.78 $\pm$ 0.05 & $0.60 \times 0.44$\\
\tableline
MM2 & \cssiete \dotfill & $37.2 \times 24.8$  & 57.96 $\pm$ 0.03  &
 5.27 $\pm$ 0.09    & 9.41 $\pm$ 0.13 & $0.66 \times 0.44$\\
    & \csdos \dotfill   & $41.2 \times 37.2$  & 57.35 $\pm$ 0.01  &
 5.19 $\pm$ 0.03    & 15.41 $\pm$ 0.07  & $0.74 \times 0.66$\\
    & \hcotrece \dotfill& $41.2 \times 33.0$  & 57.88 $\pm$ 0.02  &
 3.96 $\pm$ 0.05    & 3.52 $\pm$ 0.04 & $0.74 \times 0.60$\\
\tableline
MM3 & \tco \dotfill     & $74.2 \times 53.6$  & 58.77 $\pm$ 0.02  &
 3.01 $\pm$ 0.04    & 10.31 $\pm$ 0.11 & $1.34 \times 0.96$\\
    & \csdos \dotfill   & $86.6 \times 70.0$  & 58.73 $\pm$ 0.01  &
 3.43 $\pm$ 0.03    & 3.44 $\pm$ 0.02 & $1.54 \times 1.24$\\
\enddata
\end{deluxetable}

\clearpage

\begin{table}

\begin{center}
\caption{Derived LTE Core Parameters
\label{tbl-derivparams1}}

\begin{tabular}{llcccc}
\tableline\tableline
     &     &              & N\tablenotemark{a} & N(H$_2$)  & M$_{LTE}$ \\
Core & Transition & $\tau _{\nu}$&(cm$^{-2}$)&(cm$^{-2}$)&  (\Msun)  \\
\tableline
MM1 & \cdo      & 0.10 & $5.2 \times 10^{15}$ & $2.0 \times 10^{22}$ & 330  \\
    & \tco      & 0.75 & $3.9 \times 10^{16}$ &                    &      \\
\tableline
MM2 & \cdo      & 0.18 & $1.1 \times 10^{16}$ & $4.3 \times 10^{22}$ & 1460  \\
    & \tco      & 1.39 & $8.7 \times 10^{16}$ &                    &       \\
\tableline
\end{tabular}
\tablecomments{Velocity range for the MM1 core: 54.6--59.3 \kms.\\
$~~~~~~~~~~~~~~~~\,$Velocity range for the MM2 core: 53.4--59.3 \kms.}
\tablenotetext{a}{Column density for each molecule.}
\end{center}

\end{table}

\clearpage

\begin{table}

\begin{center}
\caption{Derived Virial Core Parameters \label{tbl-derivparams2}}

\begin{tabular}{clcccc}
\tableline\tableline
      &   & Radius\tablenotemark{a} & $\Delta V$  &
 M$_{vir}$  & n(H$_2$) \\
 Core & \multicolumn{1}{c}{Transition} & (pc)   & (\kms)     & (\Msun)    & (cm$^{-3}$)\\
\tableline
MM1 & \cssiete \dotfill  & 0.25 & 5.69 & 1700 & $3.9 \times 10^5$ \\
    & \csdos \dotfill    & 0.35 & 5.29 & 2060 & $1.7 \times 10^5$ \\
    & \hcotrece \dotfill & 0.26 & 2.75 & 410  & $8.3 \times 10^4$ \\
\tableline
MM2 & \cssiete \dotfill  & 0.27 & 5.27 & 1580 & $2.8 \times 10^5$ \\
    & \csdos \dotfill    & 0.35 & 5.19 & 1980 & $1.6 \times 10^5$ \\
    & \hcotrece \dotfill & 0.33 & 3.96 & 1090 & $1.1 \times 10^5$ \\
\tableline
MM3 & \tco \dotfill      & 0.57 & 3.01 & 1090 & $2.1 \times 10^4$ \\
    & \csdos \dotfill    & 0.69 & 3.43 & 1710 & $1.8 \times 10^4$ \\
\tableline
\end{tabular}

\tablenotetext{a}{Geometric mean of the observed semi-major and semi-minor
 axes.}


\end{center}
\end{table}

\clearpage

\begin{table}
\begin{center}
\footnotesize
\caption{Core Masses from Different Authors
\label{tbl-comparison}}

\begin{tabular}{cccccc}
\tableline\tableline
     & LTE Mass    & Virial Mass & LTE Mass &\multicolumn{2}{c}{\underline {~~~~~~~~~~~~~~~Dust Mass (1.2 mm)~~~~~~~~~~~~~~~~~}} 
           \\
Core & (This work) & (This work) & Shepherd et al. (2007) &Garay et al. (2004) & Rathborne et
al. (2006)  \\
 (1) &     (2)     &    (3)      &        (4)          &        (5)   & (6)     
            \\
\tableline
MM1 & 330  & 1130 & 75  & 550  & 1187  \\
MM2 & 1460 & 1510 & 690 & 2700 & 1284  \\
MM3 & -    & 1370 & -   & 780  & 301   \\
\tableline
\end{tabular}

\tablecomments{Masses are in \Msun.}

\end{center}

\end{table}

\clearpage

\begin{table}
\begin{center}
\caption{Core Parameters at Peak Positions
\label{tbl-derivparampeak}}
\begin{tabular}{clccc}
\tableline\tableline
      &            &              & N\tablenotemark{a} & N(H$_2$)   \\
 Core & \multicolumn{1}{c}{Transition} & $\tau _{\nu}$&(cm$^{-2}$)&(cm$^{-2}$) \\
\tableline
MM1 & \cdo      & 0.04 & $2.3 \times 10^{15}$ & $8.5 \times 10^{21}$    \\
    & \tco      & 0.30 & $1.7 \times 10^{16}$ &     \\
    & \cstyc    & 0.02 & $1.9 \times 10^{13}$ & $4.3 \times 10^{22}$  \\
    & \csdos    & 0.41 & $4.3 \times 10^{14}$ &     \\
\tableline
MM2 & \cdo      & 0.37 & $2.3 \times 10^{16}$ & $8.5 \times 10^{22}$    \\
    & \tco      & 2.80 & $1.7 \times 10^{17}$ &     \\
    & \cstyc    & 0.03 & $3.5 \times 10^{13}$ & $8.0 \times 10^{22}$  \\
    & \csdos    & 0.69 & $8.0 \times 10^{14}$ &     \\
\tableline
\end{tabular}
\tablecomments{Velocity range for the MM1 core: 54.0--59.3 \kms.\\
$~~~~~~~~~~~~~~~~\,$Velocity range for the MM2 core: 53.5--59.2 \kms.}
\tablenotetext{a}{Column density for each molecule.}
\end{center}

\end{table}

\clearpage

\begin{deluxetable}{llccccc}
\tabletypesize{\small}
\tablecaption{Derived Parameters of Molecular Outflows
\label{tbl-derivparams-outflows}}
\tablewidth{0pt}

\tablehead{
\colhead{} & \colhead{} & \colhead{M$_{\mathrm{lobe}}$}  &
 \colhead{M$_{\mathrm{hidden}}$} & \colhead{M$_{\mathrm{total}}$}
 & \colhead{\underline{~P$^{\mathrm{min}}$~~~~~~~~~P$^{\mathrm{max}}$}}
 & \colhead{~~~~\underline{~E$_{\mathrm{k}}^{\mathrm{min}}$~~~~~~~~
E$_{\mathrm{k}}^{\mathrm{max}}$}}\\
Core & Lobe & (\Msun)  & (\Msun)  & (\Msun) & (\Msun\ \kms)
 &~~~ (\Msun\ km$^2$ s$^{-2}$)\\
}
\startdata
\\[-0.9cm]
\tableline
MM1 & Blueshifted  & 3  & 3  & 6  & $2.8 \times 10^{1}~~~1.1 \times 10^{2}$ &$~~~9.2 \times 10^{1}~~~9.2 \times 10^{2}$\\
    & Redshifted   & 13 & 6  & 19 & $9.5 \times 10^{1}~~~3.3 \times 10^{2}$ &$~~~3.5 \times 10^{2}~~~2.9 \times 10^{3}$\\
    & Total        & 16 & 9  & 25 & $1.2 \times 10^{2}~~~4.4 \times 10^{2}$ &$~~~4.4 \times 10^{2}~~~3.8 \times 10^{3}$\\
\tableline
MM2 & Blueshifted  & 15 & 21 & 36 & $1.8 \times 10^{2}~~~5.5 \times 10^{2}$ &$~~~6.4 \times 10^{2}~~~4.2 \times 10^{3}$\\
    & Redshifted   & 32 & 12 & 44 & $2.2 \times 10^{2}~~~6.7 \times 10^{2}$ &$~~~7.8 \times 10^{2}~~~5.2 \times 10^{3}$\\
    & Total        & 47 & 33 & 80 & $4.0 \times 10^{2}~~~1.2 \times 10^{3}$ &$~~~1.4 \times 10^{3}~~~9.4 \times 10^{3}$\\
\tableline
MM3 & Blueshifted  & 10 & 14 & 24 & $6.8 \times 10^{1}~~~3.5 \times 10^{2}$ &$~~~1.8 \times 10^{2}~~~2.5 \times 10^{3}$\\
    & Redshifted   & 8  & 9  & 17 & $4.9 \times 10^{1}~~~2.4 \times 10^{2}$ &$~~~1.1 \times 10^{2}~~~1.8 \times 10^{3}$\\
    & Total        & 18 & 23 & 41 & $1.2 \times 10^{2}~~~5.9 \times 10^{2}$ &$~~~2.9 \times 10^{2}~~~4.3 \times 10^{3}$\\
\enddata
\end{deluxetable}

\end{document}